\documentclass[3p,final,a4paper,authoryear,twocolumn,times]{elsarticle}
\usepackage[hyphens]{url} 

\graphicspath{{./figs/}}

\DeclareGraphicsExtensions{.pdf,.png}

\usepackage[colorlinks,citecolor=blue,linkcolor=blue,urlcolor=blue,pdftex]{hyperref}

\usepackage{journalabbrevs}
\usepackage{doi}  

\newcommand{\degr}{\circ}               

\newcommand{\Gcs}{G_\mathrm{cs}}  
\newcommand{\Gnet}{G^\mathrm{net}}  
\newcommand{\Gnetcs}{G^\mathrm{net}_\mathrm{cs}}  
\newcommand{\sclearness}{k_\mathrm{S}} 

\newcommand{\STC}{\mathrm{STC}}
\newcommand{\etarel}{\eta_\mathrm{rel}}
\newcommand{\Tmod}{T_\mathrm{mod}}

\renewcommand{\Pr}{P_\mathrm{r}}  
\newcommand{\Uci}{U_\mathrm{ci}}  
\newcommand{\Uco}{U_\mathrm{co}}  
\newcommand{\Ur}{U_\mathrm{r}}    

\newcommand{\Pdensu}{\hat{P}_U}     

\newcommand{\CFu}{\mathrm{CF}_U}  
\newcommand{\CFg}{\mathrm{CF}_G}  

\newcommand{\Ptot}{P_\mathrm{tot}} 

\newcommand{\degC}{^\degr\mathrm{C}}  
\newcommand{\windunit}{\mathrm{m\,s}^{-1}}
\newcommand{\hour}{\mathrm{h}}

\newcommand{\GW}{\mathrm{GW}}

\newcommand{\Wpersqm}{\mathrm{W}\,\mathrm{m}^{-2}}

\newcommand{\densunit}{\mathrm{kg}\,\mathrm{m}^{-3}}

\begin{document}
\begin{frontmatter}

\title{The climatological relationships between wind and solar energy supply in Britain}
\author{Philip E. Bett\corref{corpb}}
\ead{philip.bett@metoffice.gov.uk}
\author{Hazel E. Thornton}

\address{Met Office Hadley Centre, FitzRoy Road, Exeter EX1 3PB, United Kingdom}
\cortext[corpb]{Corresponding author}

\begin{abstract}
We use reanalysis data to investigate the daily co-variability of wind and solar irradiance in Britain, and its implications for renewable energy supply balancing.
The joint distribution of daily-mean wind speeds and irradiances shows that irradiance has a much stronger seasonal cycle than wind, due to the rotational tilt of the Earth.  
Irradiance is weakly anticorrelated with wind speed throughout the year ($-0.4 \lesssim \rho\lesssim -0.2$): there is a weak tendency for windy days to be cloudier. This is particularly true in Atlantic-facing regions (western Scotland, south-west England).
The east coast of Britain has the weakest anticorrelation, particularly in winter, primarily associated with a relative increase in the frequency of clear-but-windy days. 
We also consider the variability in total power output from  onshore wind turbines and solar photovoltaic panels.  In all months, daily variability in total power is always reduced by incorporating solar capacity.   The scenario with the least seasonal variability is approximately $70\%$-solar to $30\%$-wind. 
This work emphasises the importance of considering the full distribution of daily behaviour rather than relying on long-term average relationships or correlations.  In particular, the anticorrelation between wind and solar power in Britain cannot solely be relied upon to produce  a well-balanced energy supply.


\copyright\ Crown Copyright 2015, Met Office.
\end{abstract}



\end{frontmatter}

\section{Introduction}\label{s:intro}
It is well known that the British Isles are in an ideal geographic situation for exploiting wind energy, and promoting wind energy has been central to  UK government policy on low-carbon energy \citep[e.g. the original version of the Renewable Energy Roadmap, ][]{decc2011roadmap}.  However, electricity generation from solar photovoltaic panels (hereafter, solar PV\footnote{Note that we focus exclusivly on electrical energy generation in this paper, and in Britain, photovoltaic panels are by far the dominant mode of electricity generation that uses solar energy directly.}) has seen huge growth in recent years, driven largely by global economic factors \citep{Bazilian2013Reconsidering, Candelise2013Dynamics}.  Reductions in the cost of PV panels have  helped to make large-scale use of solar PV in the UK financially viable, resulting in corresponding adjustments to government policy (e.g. the update to the Renewable Energy Roadmap, \citealt{decc2013roadmap}, and the Solar PV strategy \citealt{decc2013solar1, decc2014solar2}).  

Both wind and solar power output are highly variable \citep{srrench3solar, srrench7wind, Watson2013Quantifying}.   This covers weather variations on timescales of  minutes and hours, through to days and seasons, and even to long-period climate variations occurring over years and decades, linked to climate indices such as the North Atlantic Oscillation \citep[NAO,][]{Hurrell2003Overview, Scaife2008European, Colantuono2014Signature}.  However, while the variability of both is ultimately driven by the rotation of the Earth under the Sun, wind speed and irradiance exhibit different variability characteristics.  It has become increasingly important therefore to understand the relationship between energy supplied by wind and by solar PV, and the extent to which  variability in one source can help to balance out the variability in the other\footnote{Prior to the recent solar energy boom, work on the co-variability of renewable energy sources in the UK had focused on the relationship between wind and marine  energy sources \citep[wave and tidal power, e.g.][]{carbontrust2006}.  While this is now less of an immediate priority, it may become an important issue again in future.}.  This has important practical implications in terms of the need for energy storage and/or back-up capacity (e.g. from pumped storage, gas or nuclear power stations), and for the operational requirements of electricity networks.

There have been many different studies looking into these issues, with a variety of different aims, regions of interest and methodological approaches.  
\cite{Coker2013Measuring} focused on a single area in the Bristol Channel (south-west Britain), using observational records for the year 2006.  They demonstrated a range of different statistical approaches to assessing the variability of the wind, solar and tidal current energy resources in that region, on timescales of half-hours to the full year's seasonal cycle.   \cite{SantosAlamillos2012Analysis} used canonical correlation analysis to find the optimal spatial distribution of wind and solar farms across the southern Iberian Peninsula to minimise the resulting net variability.  However,  \cite{Monforti2014Assessing} found that the specific locations of generation sites made very little difference to balancing between hypothetical wind and solar supply in Italy. 
\cite{Heide2010Seasonal, Heide2011Reduced} modelled the energy storage and balancing requirements for Europe under a hypothetical high-renewable scenario, balancing wind and solar supply against demand using data spanning 2000--2007.  While their modelling is more detailed, and makes more assumptions than the work we present here, some of their conclusions are very general and important: the optimal balance of wind and solar supply to match  demand (over Europe, circa 2007) requires large amounts of storage and/or balancing supply.  They show that this can be reduced by allowing excess supply (i.e. frequent instances of supply exceeding demand), and the amounts involved affect the optimal mix between wind and solar power.  Whether one considers hourly or daily variability also has a strong impact on the relationships.  Other studies have also looked into finding the `optimal' combination of wind and solar power, for different regions and using data spanning different periods \citep{Lund2006Largescale, Widen2011Correlations, Sousa2013Optimal}, and studies generally find that incorporating both renewable sources acts to reduce the net variability in power supply, reducing the need for reserves (e.g. \citealt{Halamay2011Reserve, Hoicka2011Solar, Liu2013Analysis} in addition to those already mentioned; see also the recent review of \citealt{Widen2015Variability}).

Our present study differs from these in several key respects. Firstly, we are interested primarily in the wind--solar co-variability across Great Britain (GB) as a whole; many of the studies above use data from a limited number of specific sites.  We effectively assume that electricity networks will be able to redistribute power sufficiently to work around local imbalances.  Secondly, we are looking to avoid detailed modelling of the GB power system itself, such as details of the locations of wind and solar farms, their capacities, network connectivity, available storage etc. This information is likely to change significantly from year to year, in terms of total capacity, its partitioning between different energy sources, its geographic distribution etc., and this could have a significant impact \citep{Drew2015Impact}, limiting the applicability of our results.  Our study intends to focus on more general climatological features, based on historical data from recent decades.

Accordingly, we are also not considering electricity demand.  Ultimately, the importance of balancing wind and solar power lies in  whether or not they can together help match demand -- i.e. it doesn't matter if wind power is low, if demand is also low at the time.  However, modelling demand, including separating its socioeconomic and meteorological dependencies, requires significant attention in itself, and is beyond the scope of this paper.  Furthermore, the demand profile of Britain is likely to be significantly different in the future \citep[e.g.][]{Drysdale2014Flexible}, which adds extra uncertainties to such work.  It is useful to know what the relative behaviour is between potential wind and solar power output, as this forms the general, theoretical basis for subsequent practical applications that use particular demand/generation scenarios.

Finally, unlike most of the studies referenced above, we are focusing on the impacts of \emph{climate} variability on wind and solar energy supply: we consider monthly and seasonal  variability based on many years of daily data.  We are not considering either sub-daily or interannual/decadal variability.  Both are important, for example for understanding the frequency of ramping events \citep[e.g.][]{Cannon2015Using}, or the likely output over the lifetime of a wind farm \citep{Bett2013European, 2014arXiv1409.5359B, KirchnerBossi2014Longterm} or solar installation \citep{Allen2013Evaluation, Colantuono2014Signature}, but are outside the scope of this paper.  In the present study we are interested in understanding the distribution of possible wind--irradiance states, and treat different years as samples from an underlying climatological distribution.

This paper proceeds as follows.  In section~\ref{s:data} we describe the data we have used, and our analysis techniques.  Our results on the joint distributions of wind and irradiance are discussed in section~\ref{s:results}.  We discuss the impact of the wind--irradiance distribution on the resulting total power variability for different scenarios in section~\ref{s:balancing}.  Our conclusions are presented in section~\ref{s:concs}.

\section{Data and methods}\label{s:data}
This study uses the ECMWF\footnote{The European Centre for Medium-range Weather Forecasting.} Re-Analysis Interim data set (ERA-Interim), which is described in full in \cite{Dee2011ERAInterim} and \cite{Berrisford2011ERAInterim}.  We have obtained ERA-Interim data covering 1979--2013, at $0.75^\degr$ spatial resolution.

Despite being based on assimilations of vast amounts of observational data, from ground stations as well as satellites, the low spatial resolution of the ERA-Interim data means that it should be treated with caution when comparing its results to observations.  We use ERA-Interim because its wind speeds, irradiances, temperatures etc. are produced from the same physical model, constrained by observations. This means that they are physically consistent at any given time step, and at a consistent spatial scale.  This is a distinct advantage over using a mixture of data sources, such as  reanalysis in conjunction with satellite-based or station-based observations.  We reiterate, the goal is to assess the co-variability of wind and solar resources at the GB scale, \emph{not} to produce a detailed, accurate description of the available resource.

A recent study by \cite{Boilley2015Comparison} showed the deficiencies in using reanalysis data for estimates of irradiance, compared to satellite-based data.  They found that reanalyses tend to have too many clear-sky days compared to observations, although in the particular case of ERA-Interim this is countered somewhat by also having many cloudy days that were observed to be clear.   A substantial amount of the true variability  in irradiance \emph{at a site} is not captured in reanalysis data.  The difficulty in our case, motivating our decision not to use satellite data, is in finding comparable high quality wind speed data that we can use in our co-variability assessment.  \cite{Kubik2013Exploring} assessed the use of reanalysis data for regional wind assessments, and found that its benefits, such as its continuous nature over a long time period,  outweighed the disadvantages from low resolution, when used with care.

Using ERA-Interim, we have taken the daily ($24\,\hour$) means of the 6-hourly wind speed fields at model level 58, which corresponds to a height of roughly 60\,m above ground level.  We denote these wind speeds by $U$.

Daily-mean downwelling shortwave irradiance\footnote{Irradiance is the radiative energy flowing through a unit area per unit time; cf. irradiation or insolation, which is the total radiative energy per unit area, integrated over a given time.} at the surface\footnote{For brevity, all irradiance/irradiation fields in this paper should be interpreted as referring to downwelling shortwave radiation at the surface, unless otherwise noted.} is not directly available from the ERA-Interim data archive, and has to be calculated from the 3-hourly forecast fields for accumulated irradiation.  The resulting fields are daily-mean downwelling total irradiances at the surface, for a horizontal plane.  Since ERA-Interim does not track the direct and diffuse radiation components separately, we are not able to calculate the irradiance falling on a tilted surface.  Because total irradiance (direct $+$ diffuse) is often termed `global' irradiance, we denote it by $G$.

Much of the local variability in solar irradiance is directly due to the tilt of the Earth's rotational axis with respect to its orbit, known as its obliquity.\footnote{The seasonal cycle in wind speeds can be traced back to the same factors of course, but much less directly.} This causes variation in both the total hours of daylight and the overall intensity of the incident radiation (the amount per unit area).  These `astronomical' factors have two key features: they are entirely predictable, and they dominate the seasonal variability of irradiance (this can be seen in later figures).  However, while they are therefore critical to the amount and seasonal variability of power generation from solar PV, they also mask any direct physical relationship with wind speed.

It is useful therefore to be able to factor out the obliquity component of irradiance variability, and consider how the remainder varies with wind speed.  To do this, we compare $G$ with the daily-mean downwelling clear-sky (cloud-free) irradiance, $\Gcs$.    This field is not available in the ERA-Interim archive. Instead, we obtained the daily-mean \emph{net}  irradiances (i.e. from downwelling minus upwelling shortwave radiation) for all sky conditions $\Gnet$, and clear-sky conditions $\Gnetcs$.  We can then assume the ratios of these irradiances are the same; we shall refer to this quantity as the \emph{surface clearness}:
\begin{equation}\label{e:sclearness}
  \sclearness := \frac{G}{\Gcs} = \frac{\Gnet}{\Gnetcs}.
\end{equation}
This contrasts with the traditional definition of the (total) clearness index $k_\mathrm{T}$, which is the ratio of surface irradiance to that received at the top of the atmosphere, i.e. before any kind of atmospheric absorption.  In our case however, we want to compare the irradiance received at the surface with that which  \emph{would have been received} at the surface under clear sky conditions, i.e. in the absence of the dominant governing meteorological factor, cloud cover\footnote{In reality, aerosol levels will also have an important impact on surface irradiance. However, ERA-Interim uses a monthly climatology of aerosols, so they will not have a direct daily relationship with wind speed.}.

The surface clearness ratio therefore describes what fraction of irradiance remains after being attenuated by clouds: $\sclearness=1$ corresponds to clear skies (no clouds), and lower values of $\sclearness$ imply greater attenuation by clouds.  We are still working with daily averages: winter days in Britain will have a lower mean clear-sky irradiance $\Gcs$ than summer days due to both the  reduced intensity of radiation and the reduced day length.  The daily mean irradiance $G$ will be affected by these factors \emph{and} by clouds, so scaling $G$ by $\Gcs$ to produce $\sclearness$ leaves only the variability due to cloudiness.

We also calculate the notional power output from wind turbines and solar panels based on this data.  While the models we use for doing this are relatively simple, they provide a way of relating potential wind and solar power on a fair footing, based on internally-consistent meteorology.  For clarity, the details of this modelling is given in \ref{s:powercalcs}.

In this paper, we shall be focusing on area-weighted averages of these quantities over Great Britain (GB), after applying the ERA-Interim land--sea mask.

Finally, it will often be useful to break results down into `seasons', which we take to be 3-month periods.  We make a distinction between `solar' seasons (centred on the solstices/equinoxes, so `winter' is Nov--Dec--Jan, NDJ), and the more usual `meteorological' seasons, which are offset by a month (so `winter' is Dec--Jan--Feb, DJF) due to the seasonal lag delaying the response of temperature, wind speeds etc. to the changes in irradiance arriving at the top of the atmosphere.   Since surface irradiance will largely follow the solar seasons and wind speed will follow the meteorological seasons, we will make explicit use of both of these different definitions when appropriate.


\section{Results and discussion}\label{s:results}
\subsection{The daily co-variability of wind and irradiance}\label{s:correls}
\subsubsection{Joint distributions}\label{s:jointdistros}
\begin{figure}
\centering\includegraphics[width=\columnwidth]{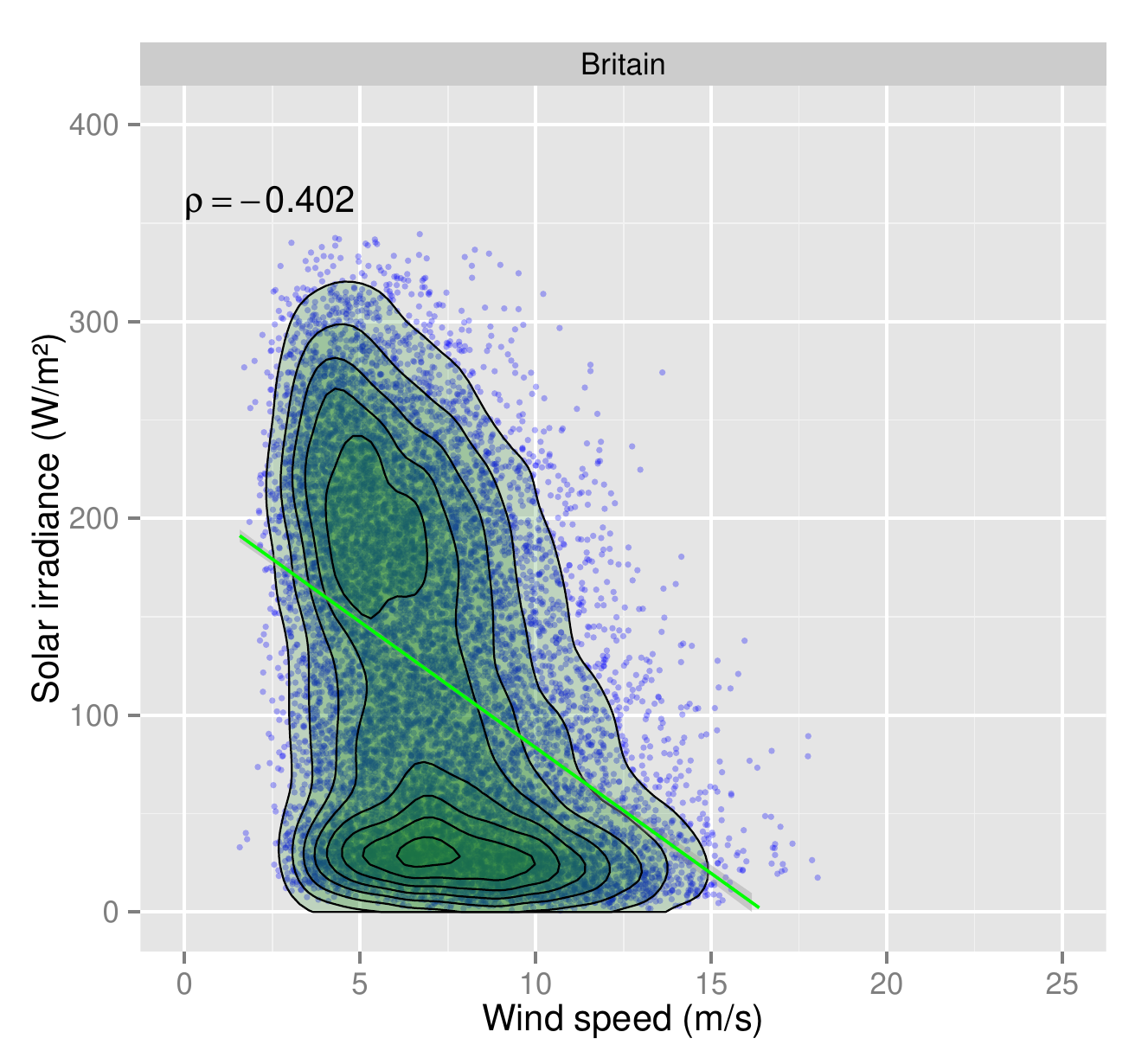} 
\centering\includegraphics[width=\columnwidth]{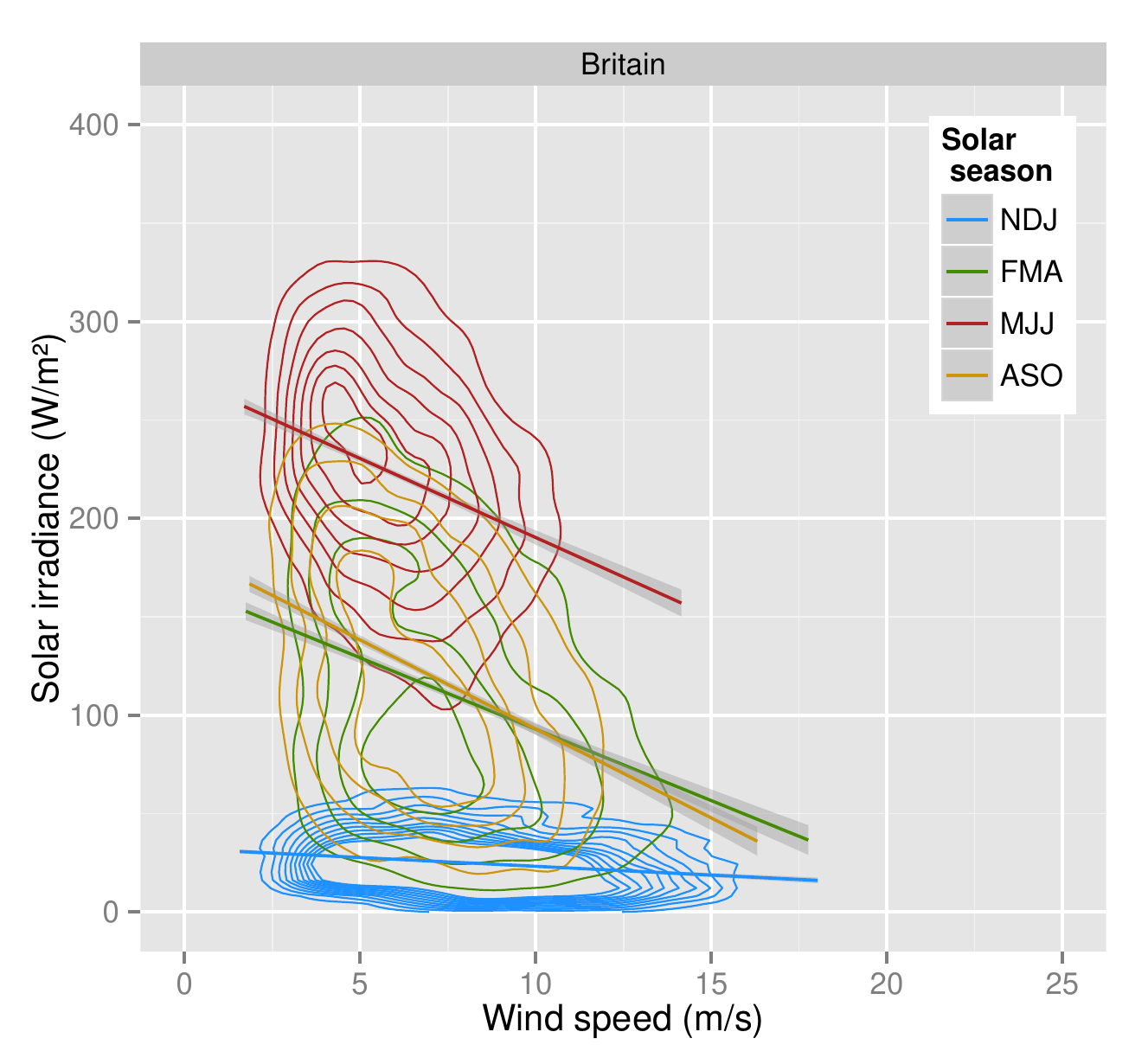}
\caption{The joint distribution between daily-mean wind speed at 60\,m, and downwelling shortwave irradiance at the surface, averaged over Britain.  \emph{Top:} Individual daily values are plotted as blue points, and the point density is shown by contours and green shading.  The linear regression line is shown in bright green, and the Pearson correlation coefficient $\rho$ is given. Contours mark densities of points between $0$ and $10^{-3}$ in steps of $10^{-4}$.  \emph{Bottom:} The same data separated into seasons, defined as indicated. Density contours are plotted between 0 and $2\times 10^{-3}$ in steps of $2\times 10^{-4}$. The linear regression lines and their confidence intervals are also plotted for each season.  }
\label{f:jointdistro}
\end{figure}

The most direct way of looking at the relationship between daily-mean, GB-average wind speeds and solar irradiances, is simply to plot their values for each day  against one another.  The resulting joint distribution is shown in Figure~\ref{f:jointdistro}. (For completeness, we show the joint distribution in terms of power generated in \ref{s:jointdistroalt}.)

While there is an overall anticorrelation between wind speed and irradiance ($\rho \simeq -0.4$), the bimodal form of the distribution means that a simple linear fit is a poor description of the relationship.  The seasonal breakdown shown in the lower panel of Figure~\ref{f:jointdistro} clearly shows the reason for the bimodality: the strong seasonal cycle in irradiance, with bright days in summer clearly separated from darker days in winter.  This differs from the seasonality seen in the wind:  the seasonal mean wind speed varies much less than its own day-to-day variability.  For example, the density peak of the wind speed distribution in summer is within the envelope of the winter wind speeds, and vice versa.

\begin{figure}
\centering\includegraphics[width=\columnwidth]{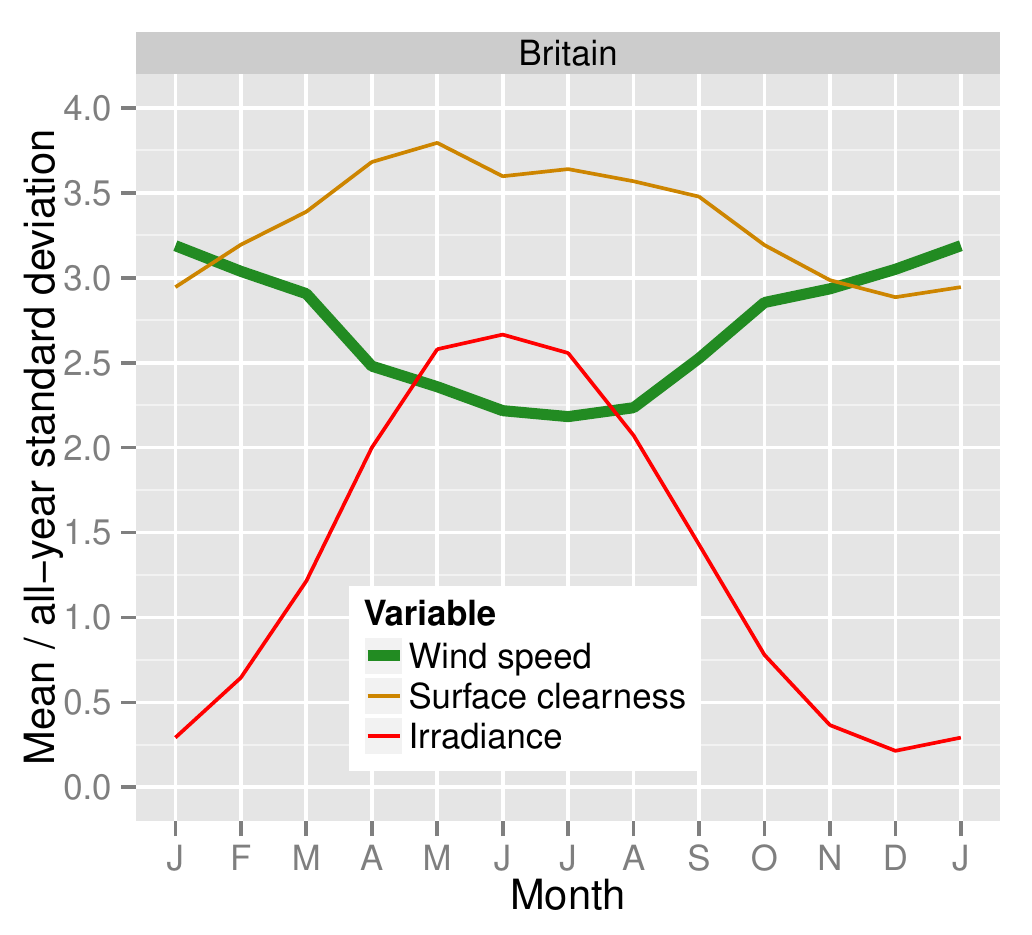}
\caption{Annual cycle of monthly mean wind speed, irradiance, and surface clearness, compared to the all-year standard deviations of their daily data.
Note that in this and subsequent similar plots, the results for January are repeated after December to show continuity of the annual cycle.
}
\label{f:monmeanoversd}
\end{figure}

This is shown explicitly in Figure~\ref{f:monmeanoversd}.  Here, we plot the monthly mean irradiance and wind speed, as fractions of their all-time standard deviations\footnote{The standard deviation of daily-mean irradiances is $85.4\,\Wpersqm$, and that of wind speeds is $2.69\,\windunit$.  The standard deviation of the surface clearness ratio is $0.187$.}, $\sigma$.  The monthly irradiances vary by nearly $2.5\sigma$.  In contrast, the monthly wind speeds vary by about $1\sigma$.

The seasonal wind variability shown in Figures~\ref{f:jointdistro} and~\ref{f:monmeanoversd} has an important message for our understanding of the wind distribution: While the highest wind speeds only occur during winter,   lower wind speeds occur throughout the year.  It is not simply the case that Britain gets low winds in summer and high winds in winter; rather, winters have more variable winds, and include higher winds that are absent in the summer.

Figure~\ref{f:monmeanoversd} also shows the monthly variability of the surface clearness ratio we defined in equation~(\ref{e:sclearness}): its monthly means vary by less than $1\sigma$, similar to wind speed, and much less than irradiance itself.

\begin{figure}
\centering\includegraphics[width=\columnwidth]{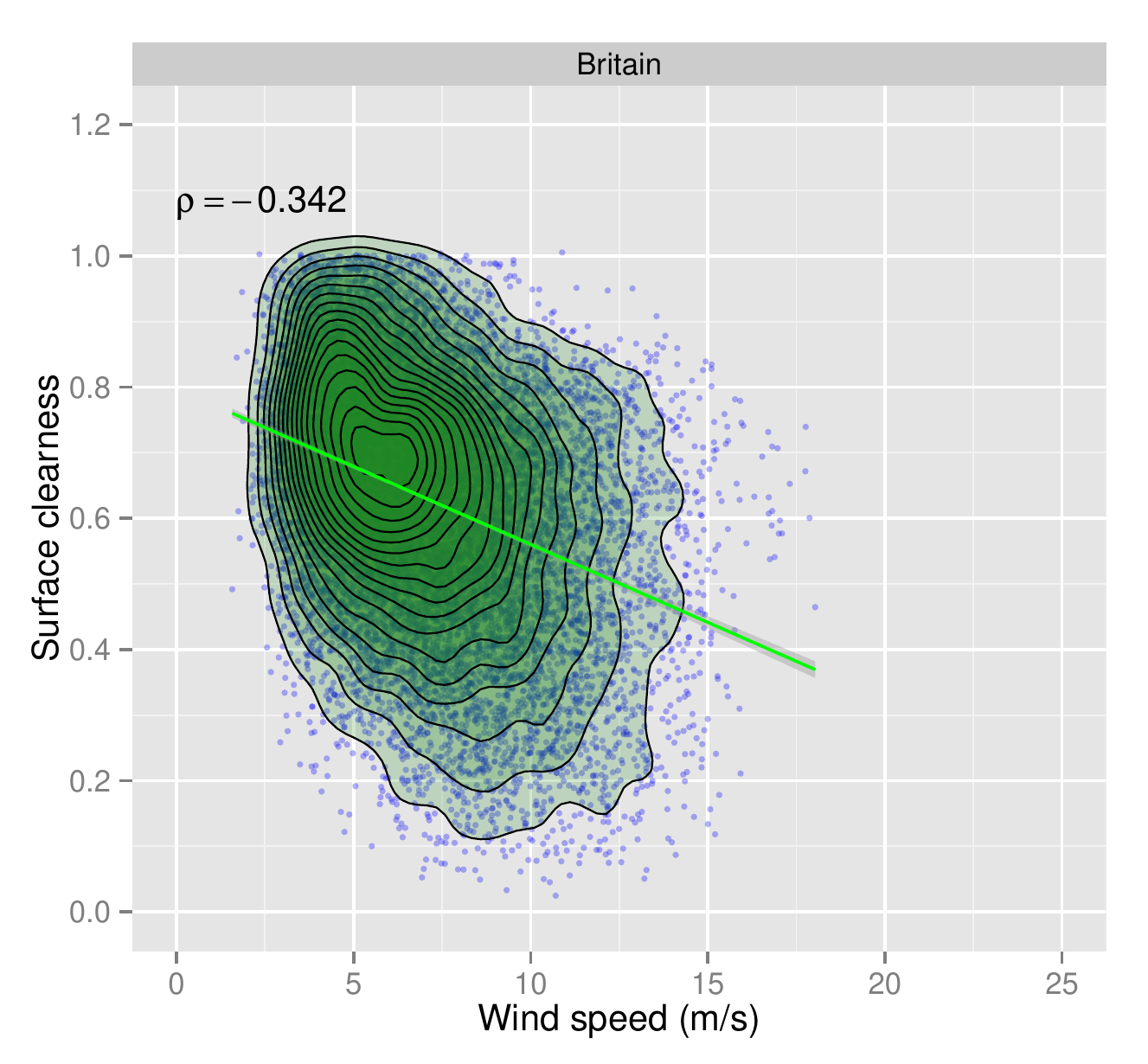} 
\centering\includegraphics[width=\columnwidth]{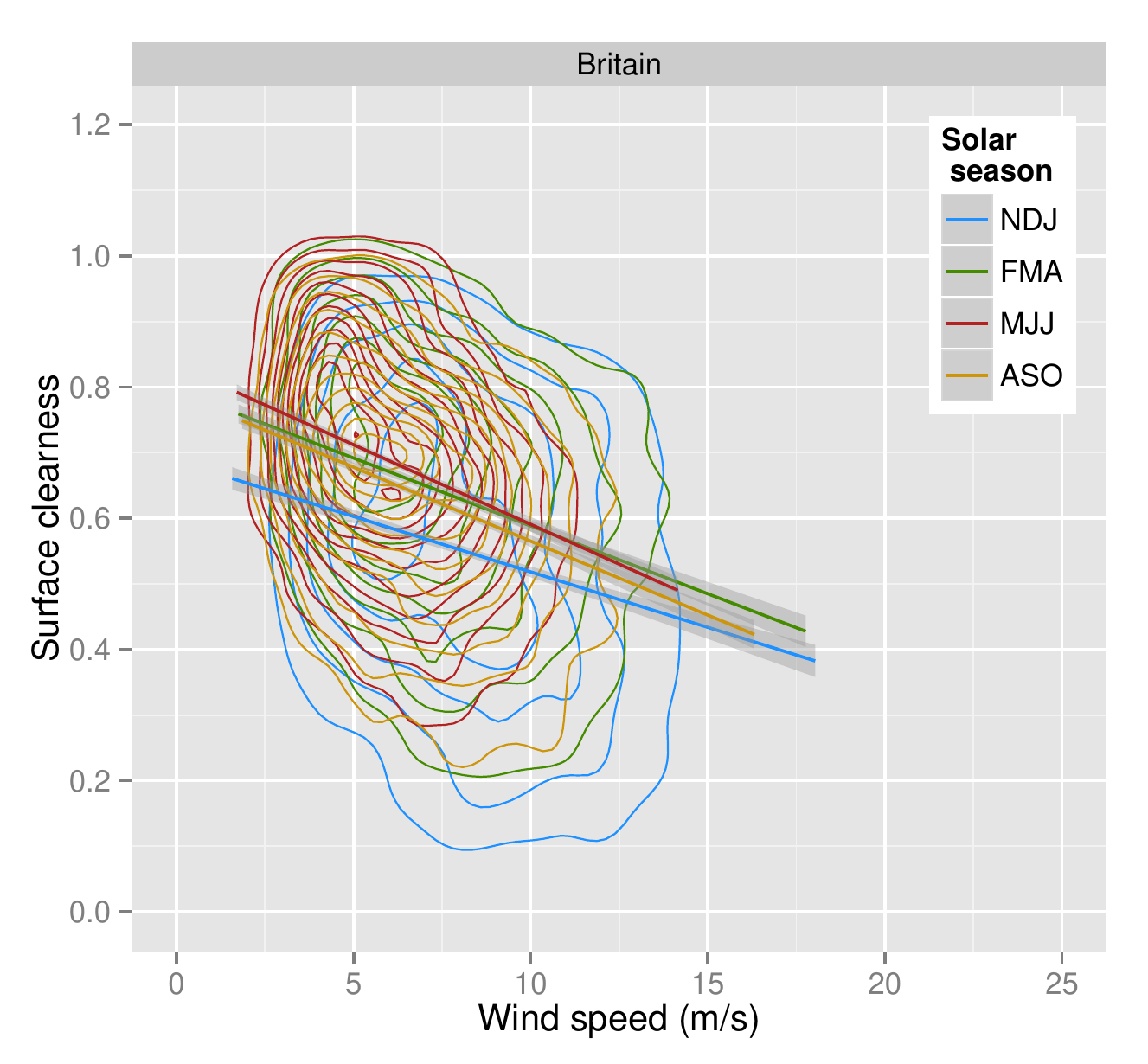}
\caption{Joint distribution of daily-mean surface clearness (equation~\ref{e:sclearness}) and wind speed, similar to Figure~\ref{f:jointdistro}.  \emph{Top:} Individual daily values are plotted, with contours marking their density between $0$ and $0.5$ in steps of $2\times 10^{-2}$, and the linear regression line overplotted. \emph{Bottom:} The same data separated into seasons, with contours between $0$ and $1$ in steps of $4\times 10^{-2}$, and with  individual linear regression lines.
}
\label{f:jointdistrowithratio}
\end{figure}

\begin{figure}
\centering\includegraphics[width=\columnwidth]{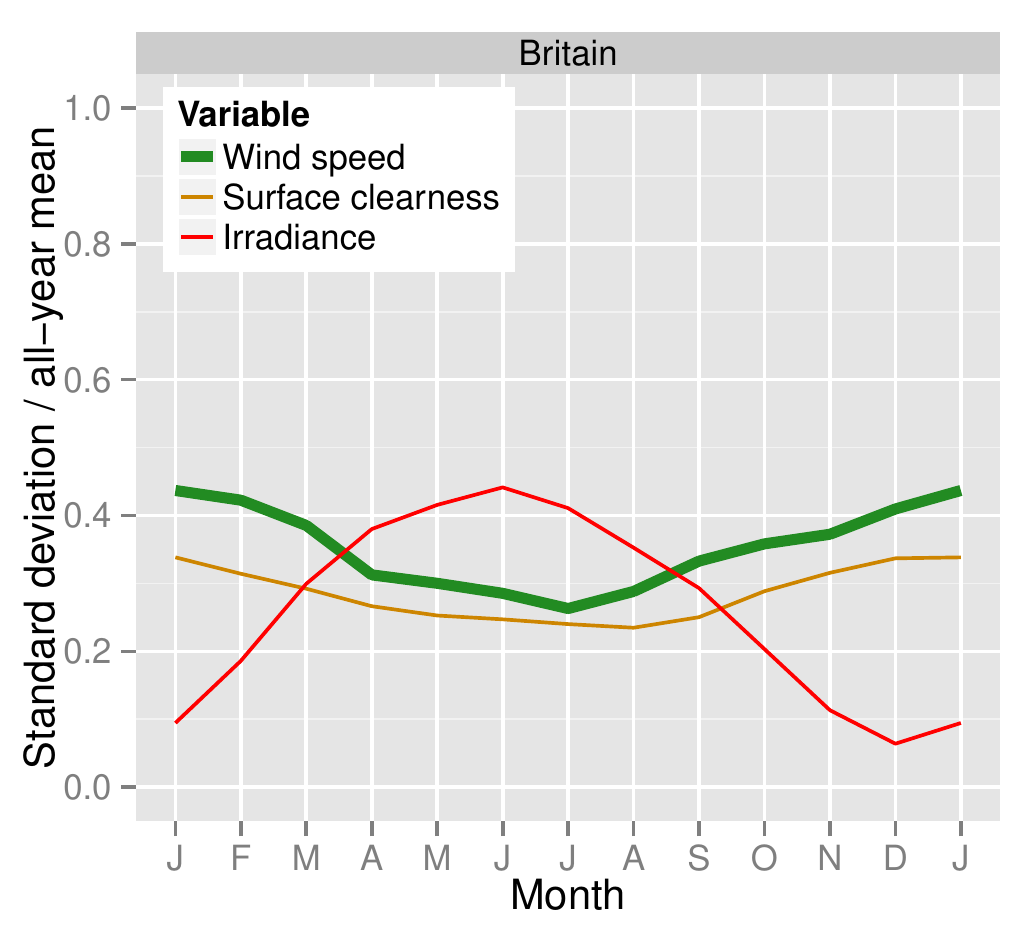}
\caption{Annual cycle of monthly variability (standard deviation of daily data) compared to the all-year mean value, for the variables indicated.
}
\label{f:monsdovermean}
\end{figure}

We show the joint distribution of surface clearness with wind speed in Figure~\ref{f:jointdistrowithratio}.  Even after the effects of obliquity have been removed, making the distributions in different seasons much more similar, the data remain weakly anticorrelated.  Furthermore,  both the wind and surface clearness exhibit greater variability in winter than summer.  We show this explicitly in Figure~\ref{f:monsdovermean}, where we plot the monthly standard deviations of the daily data, scaled by their all-time averages.  The phase reversal  between irradiance and surface clearness variability is clear, as is the overall reduction in month-to-month changes in daily variability.

\subsubsection{Seasonal variation in the correlation}
\begin{figure}
\centering
\includegraphics[width=\columnwidth]{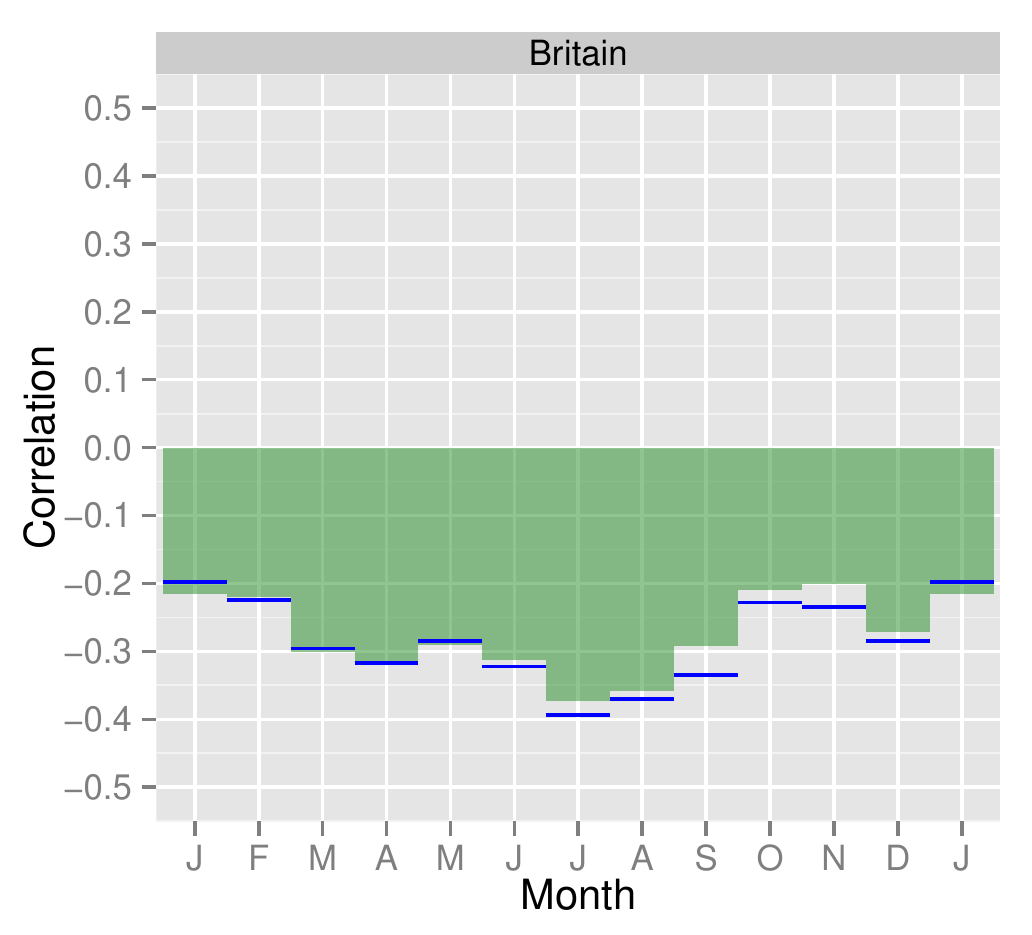} 
\caption{Monthly correlations of daily mean wind speeds with irradiances (green bars) or with surface clearness (blue lines).  
}
\label{f:monthlycorrels}
\end{figure}

As suggested by the preceding figures, the (anti-) correlation between wind speed and surface clearness is not constant over the year.  Monthly values of the correlation between wind and irradiance, and surface clearness, are shown in Figure~\ref{f:monthlycorrels}.  It is important to note that, mathematically,  the Pearson correlation is unchanged when scaling by a constant, i.e. $\mathrm{cor}(X,Y) \equiv \mathrm{cor}(X/x_0,Y)$.  In our definition of surface clearness, clear-sky irradiance is clearly not constant.  However, it exhibits much less variability from one day to the next, and between different years, than the actual irradiance.  So, for a given location and month, we expect (and find) the correlation to be largely  unchanged when replacing irradiance with surface clearness.

With both irradiance and clearness, the greatest anticorrelation with wind speed occurs in July, and the least correlated months are October/November and January.  While it is important to understand the meteorological behaviour driving this, it should not be over-interpreted: the correlation values are small throughout, and there is very little change from month to month.  

What month-to-month change in correlation there is can be explained by looking at the joint distribution in Figure~\ref{f:jointdistrowithratio}: during winter, there tends to be a wider range of clearness values at any given wind speed compared to the summer.  The more dynamic atmosphere in winter allows for more cloudier days to be included in the distribution.  This means that the value of the Pearson correlation will be closer to zero in the winter than the summer.   These results agree with  \cite{He2013Diurnal},  who showed a clear shift in wind distribution towards higher winds during cloudy conditions, especially in winter.  

\subsubsection{Spatial variation in the correlation}\label{s:spatvar}
\begin{figure}
\centering 
\includegraphics[width=\columnwidth]{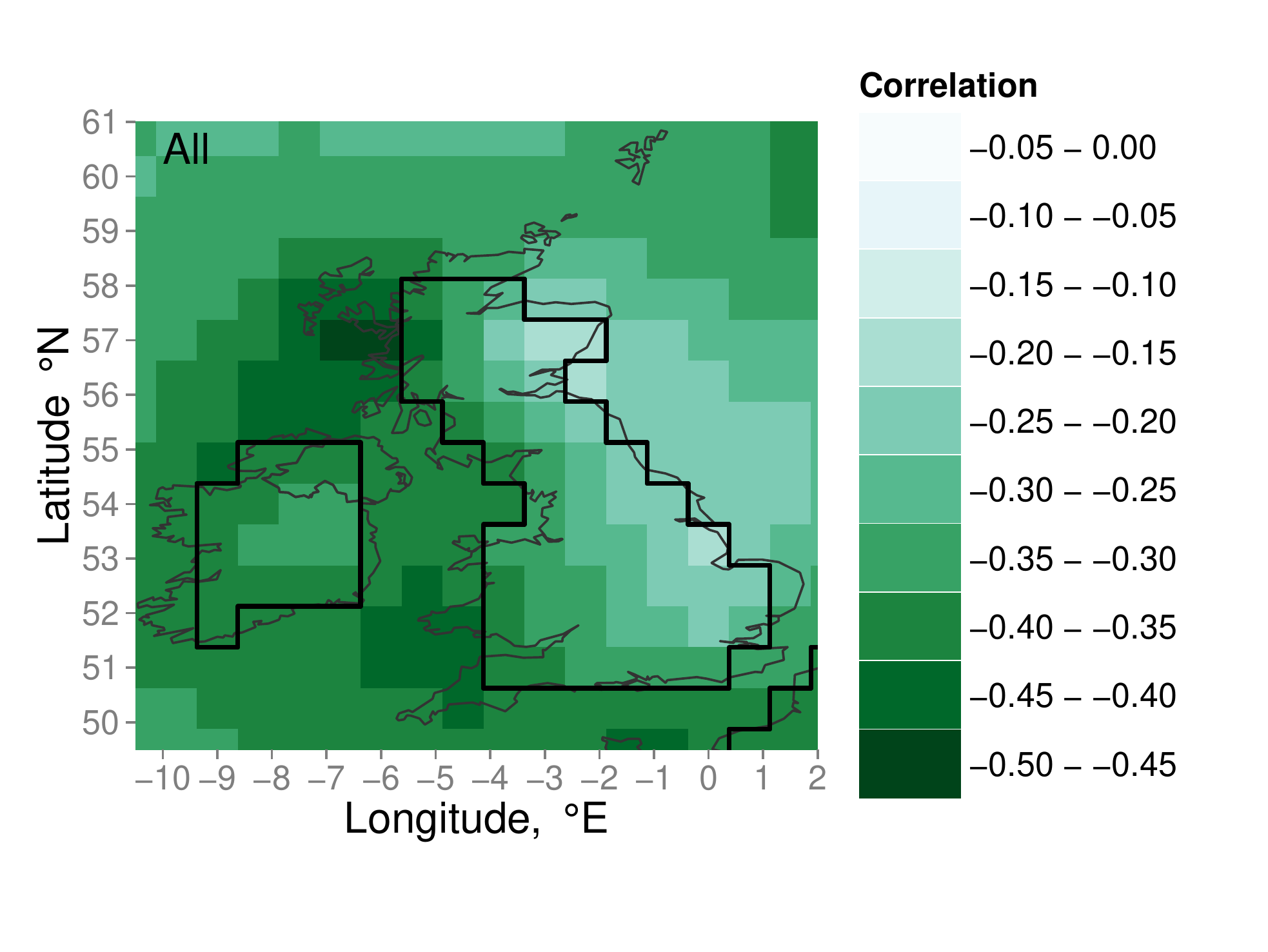}\\
\includegraphics[width=0.48\columnwidth]{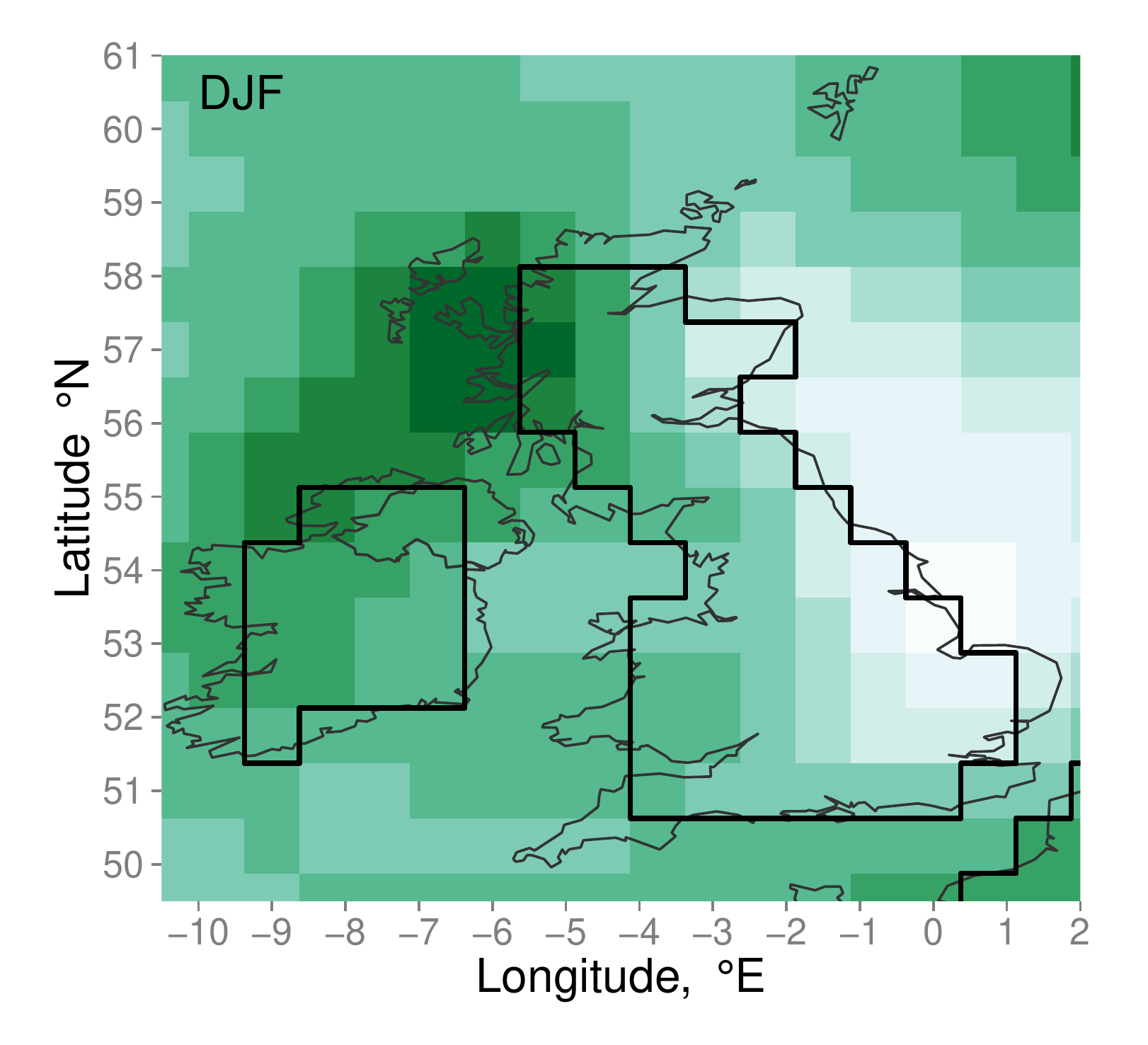}
\includegraphics[width=0.48\columnwidth]{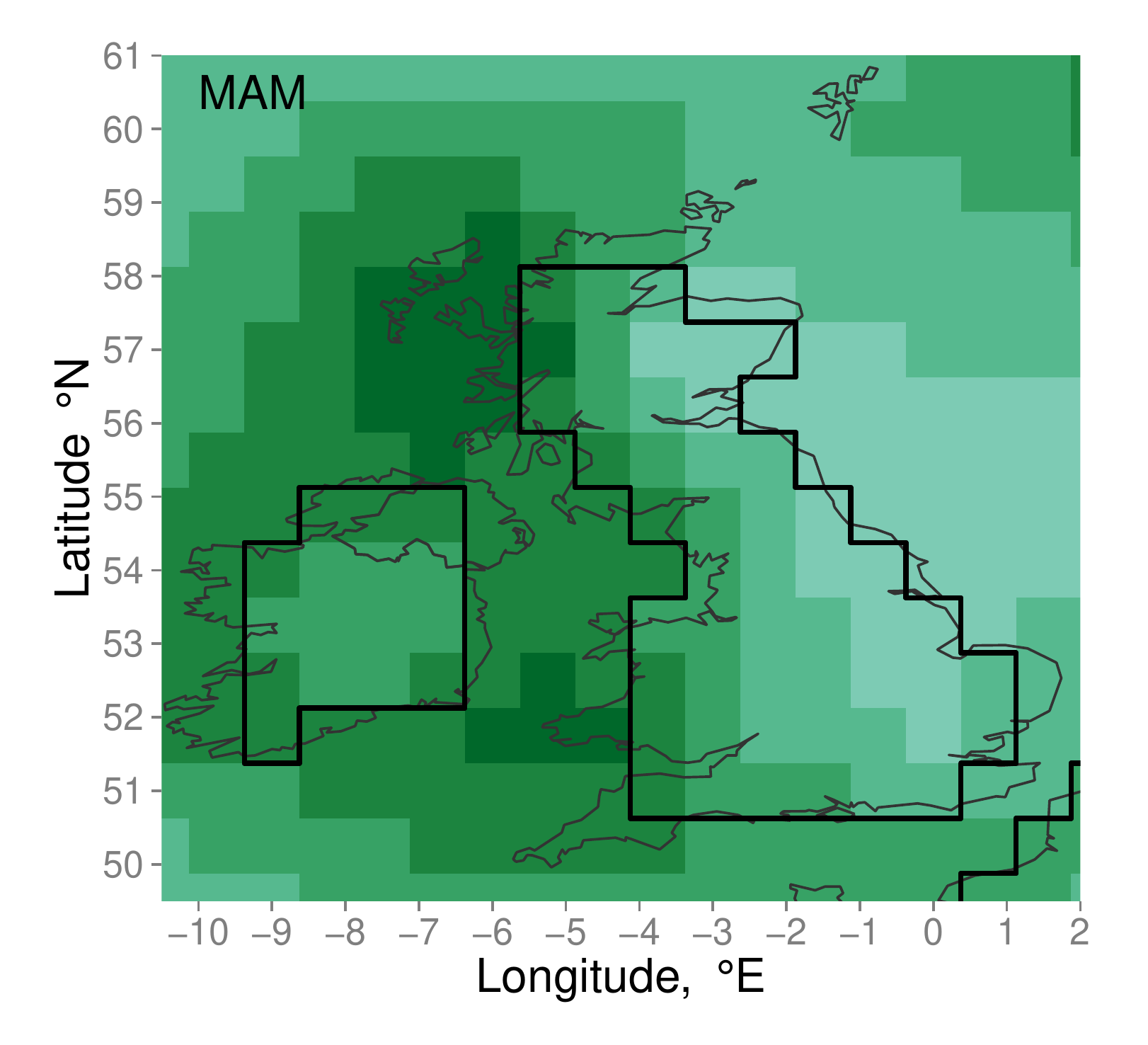}\\
\includegraphics[width=0.48\columnwidth]{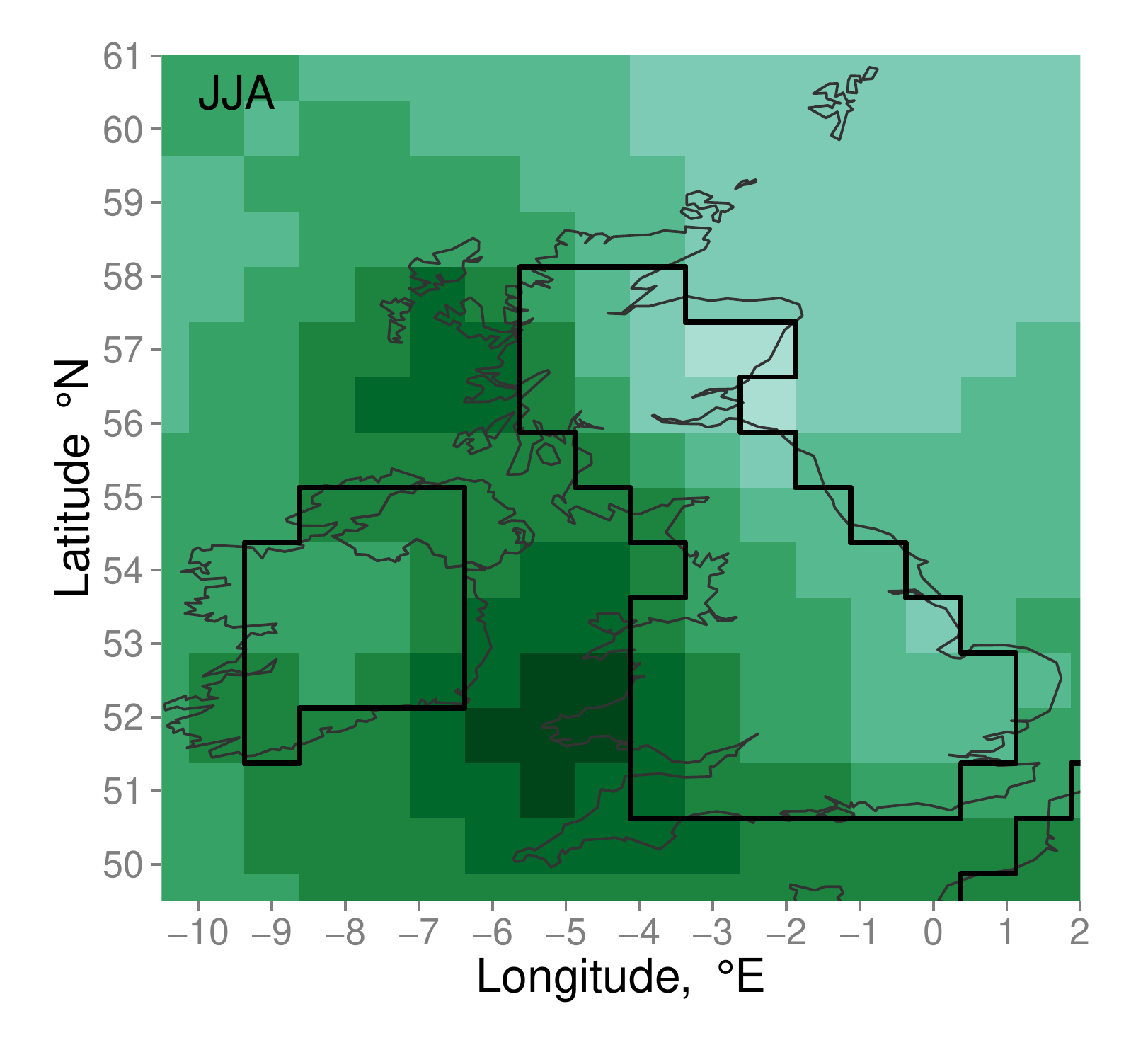}
\includegraphics[width=0.48\columnwidth]{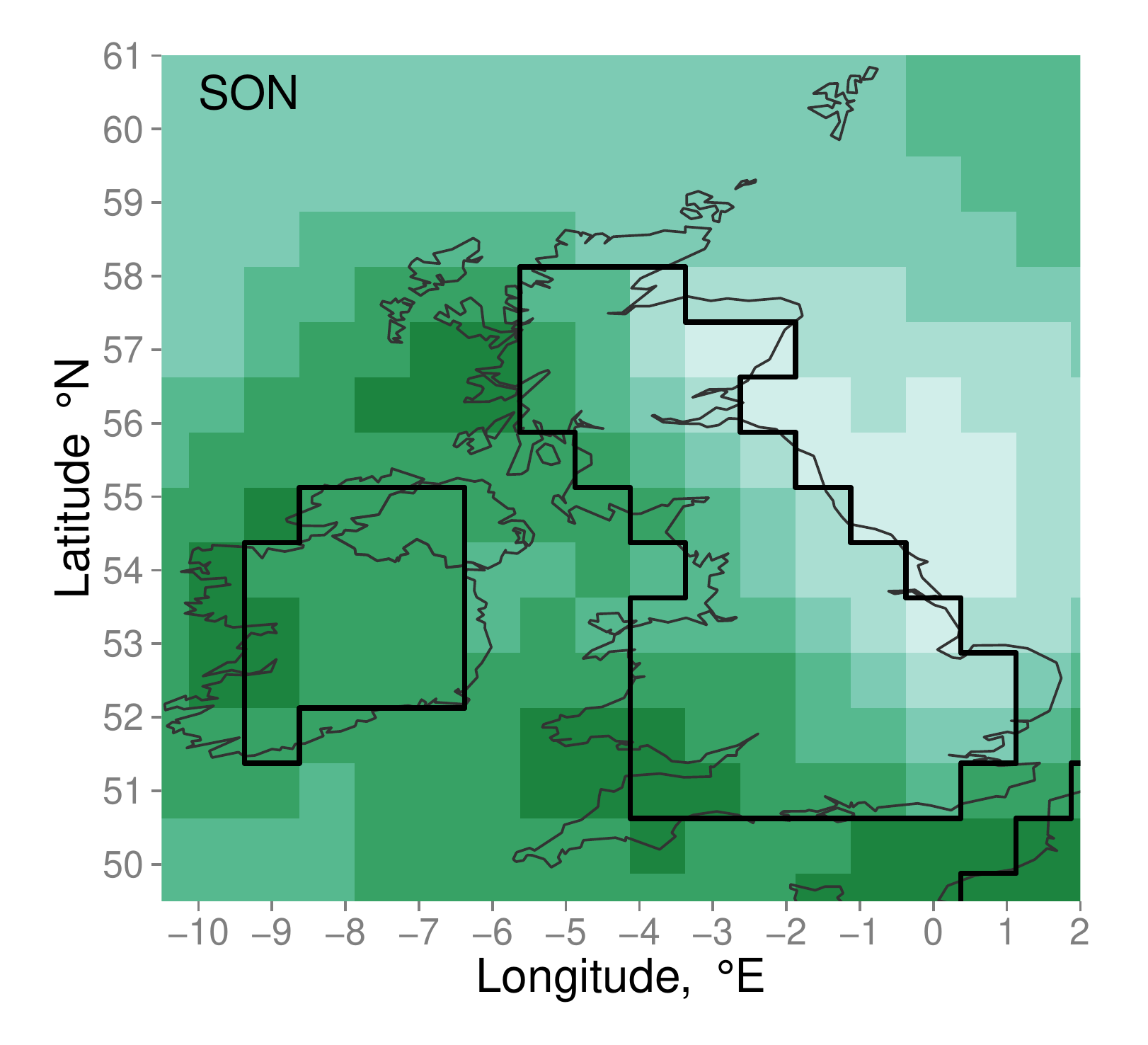}
\caption{Maps of the correlation of daily-mean wind speed with surface clearness.  The top panels show the all-year correlation, and the smaller panels show different seasons as labelled.   The same colour scale is used in all panels. 
}
\label{f:correlmap}
\end{figure}

There is also some systematic geographical variability in the correlation of wind with irradiance and surface clearness (Figure~\ref{f:correlmap}; results for irradiance are similar and shown in~\ref{s:windirradcormaps}).  In particular, the western, Atlantic-facing regions of Britain have a much stronger anticorrelation than the east coast. This is particularly noticeable when considering the surrounding seas, which we do not include in our GB-averages.  The anticorrelation off the north-west coast of Scotland reaches its greatest extent in spring, while the peak off the south Wales/south-west England coast is strongest in summer.  The correlations are much weaker over the east coast of Britain, particularly in winter.  

These results agree with those of \cite{Colantuono2014Signature}, who showed that there is an east--west gradient in the impact of the winter NAO on solar radiation, using ground station data between 1998 and 2013.  Winter wind speeds over Britain are well-known to correlate with the NAO, with strong NAO-positive winters tending to be much stormier \citep[e.g.][]{Hurrell2003Overview, Scaife2014Skillful}.    Our use of data covering 1979--2013 lets us investigate the large-scale behaviour over the whole year, with a more well-defined climatological basis.

\begin{figure}
\centering
\includegraphics[width=\columnwidth]{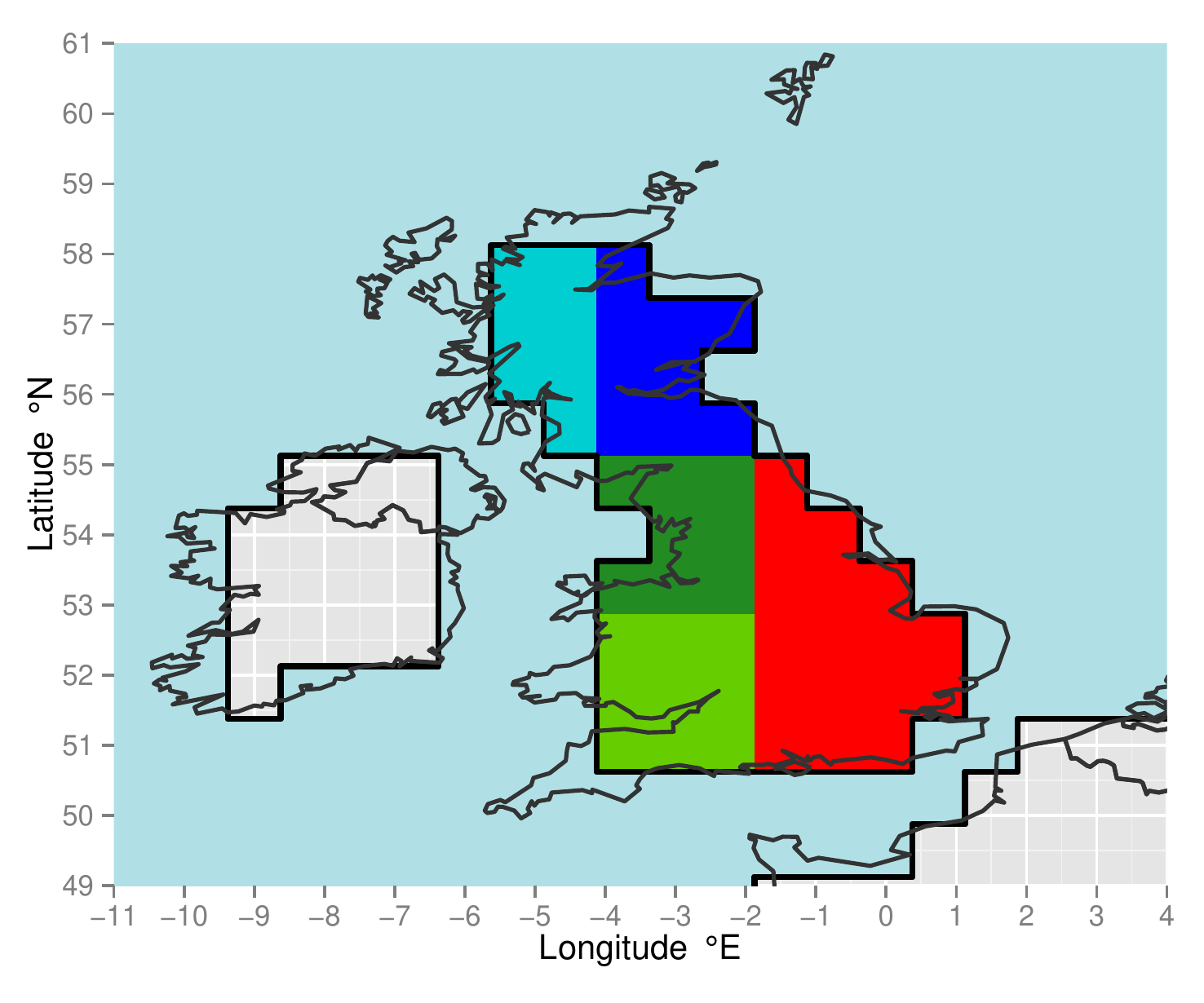} 
\caption{Definitions of regions used in Figure~\ref{f:tercilesbytercs}.  The definitions are rectangular on the lat--lon grid, but regional averages are calculated after applying the land--sea mask, as shown.
}
\label{f:regsmap}
\end{figure}

To understand what is causing the regional variations over the year, we consider how much of the time is spent in different areas of the joint windspeed--clearness distribution, over Britain as a whole, and in different sub-regions that we define in Figure~\ref{f:regsmap}.  We calculate the terciles of the surface clearness distribution and wind speed distributions separately, using all-year data.  We focus on the combinations of the upper and lower terciles of the two variables, which we label ``clear'' \& ``cloudy'' for $\sclearness$, and ``calm'' \& ``windy'' for $U$.  We calculate the fraction of days each month that is spent in each of the four combinations of these terciles.  Given the anticorrelation we have already seen, we expect the most frequently-occupied categories to be the ``clear \& calm'' and ``cloudy \& windy'' tercile combinations.  Changes in the frequencies of the opposing two combinations -- ``clear \& windy'' and ``cloudy \& calm'' -- will modulate the strength of the anticorrelation.  

\begin{figure}
\centering
\includegraphics[width=\columnwidth]{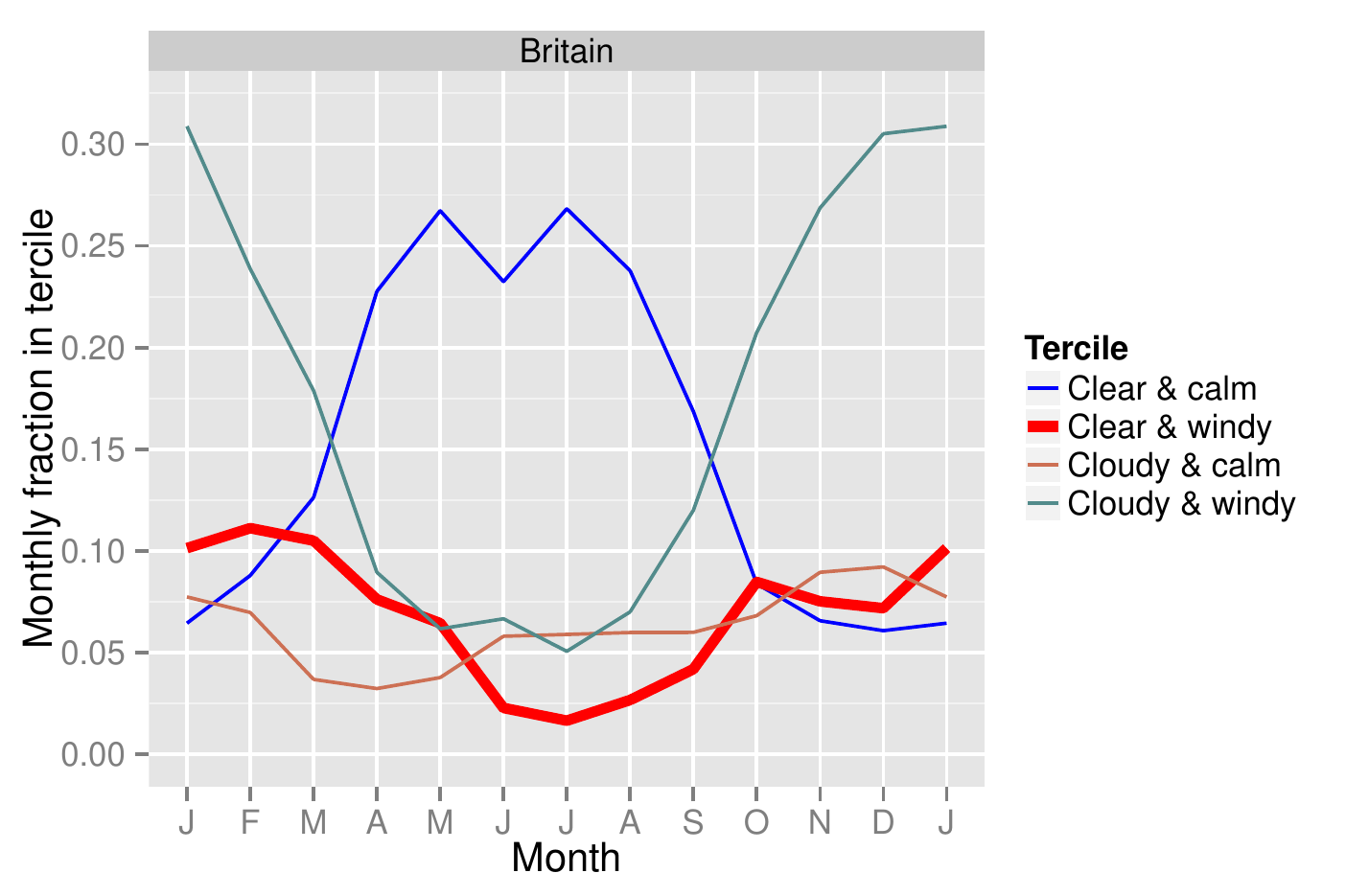}
\caption{Fractions of days each month in different combinations of terciles of the GB-average surface clearness and wind speed distributions.  ``Clear'' and ``cloudy'' correspond to the lower and upper third of the all-year clearness distribution; the lower and upper wind speed terciles are labelled  ``calm'' and ``windy''.  
}
\label{f:tercilesbyregs}
\end{figure}

Focusing initially on the seasonal variation in correlation seen in Figure~\ref{f:monthlycorrels}, we show the monthly occupation of the different tercile categories for Britain as a whole in Figure~\ref{f:tercilesbyregs}. The seasonal cycle between the dominant clear/calm and cloudy/windy tercile combinations is immediately apparent.   The cloudy/calm combination occurs at a relatively consistent low rate throughout the year.  The clear/windy combination however has a stronger seasonal cycle, occurring with greater frequency in autumn and winter.  This points to  clear-but-windy days playing a key role in reducing the level of correlation (bringing it closer to zero) in those seasons.

\begin{figure*}[t]
\centering
\includegraphics[width=\textwidth]{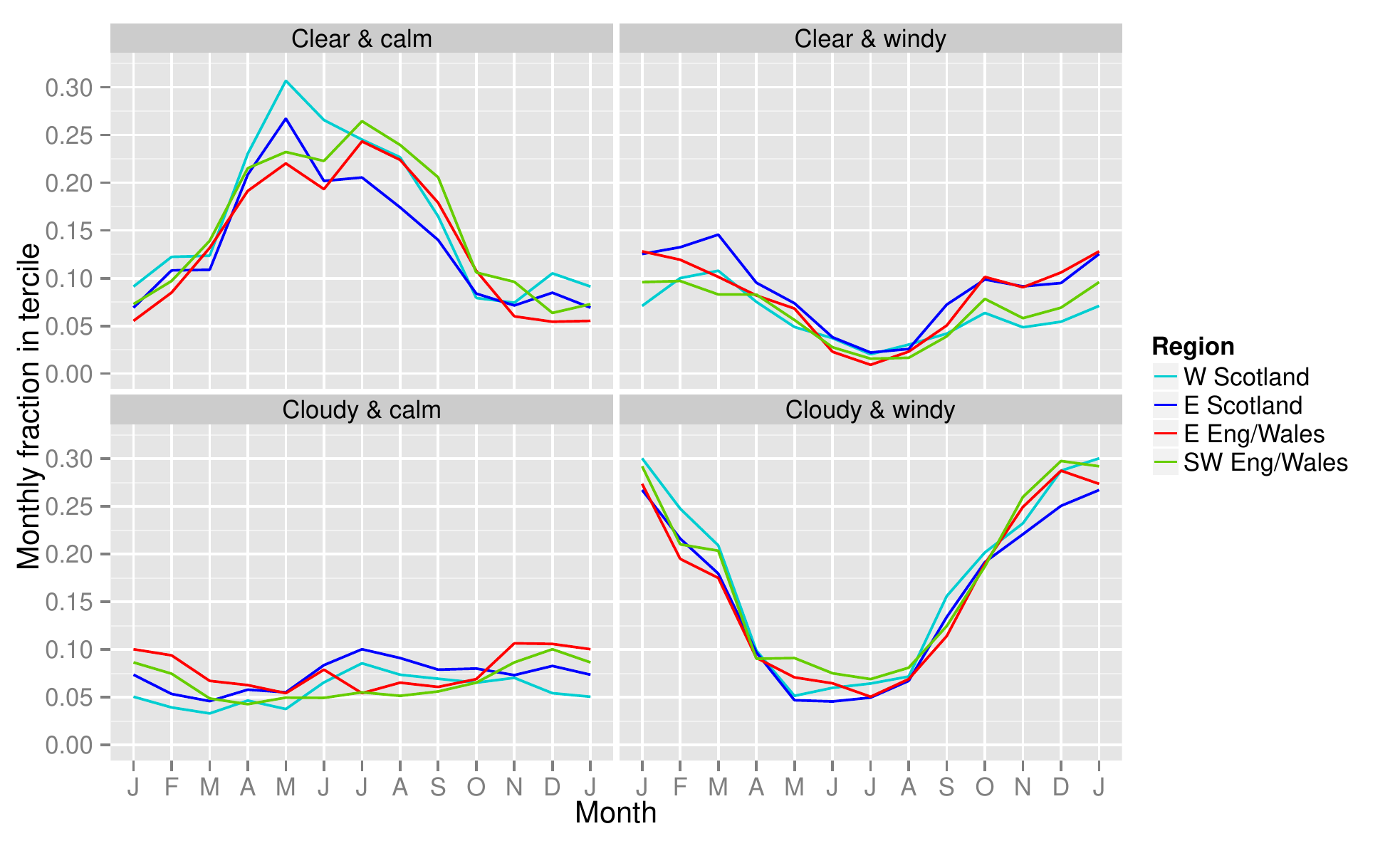}
\caption{As Figure~\ref{f:tercilesbyregs}, but grouped into the four tercile combinations, showing results for the different regions defined in Figure~\ref{f:regsmap}.  To avoid crowding, the  north-west England \& Wales region is not shown; its results are intermediate to those shown.
}
\label{f:tercilesbytercs}
\end{figure*}

We can verify this further by looking at the regional breakdown of the tercile combinations (Figure~\ref{f:tercilesbytercs}).  This shows two additional features.  Firstly, the weak cloudy/calm combination in fact shows different seasonal cycles for northern and southern regions, peaking in July for Scotland, but autumn/winter for England/Wales.

On the other hand, the clear/windy combination provides a means to differentiate between eastern and western regions, following the east/west distinction we saw in the correlation maps (Figure~\ref{f:correlmap}).  The two eastern regions, which have the weakest correlations, have a higher frequency of clear/windy days in winter than the  two Atlantic-facing regions (western Scotland and south-west England \& Wales).

\begin{figure*}
\centering
\includegraphics[width=0.49\textwidth]{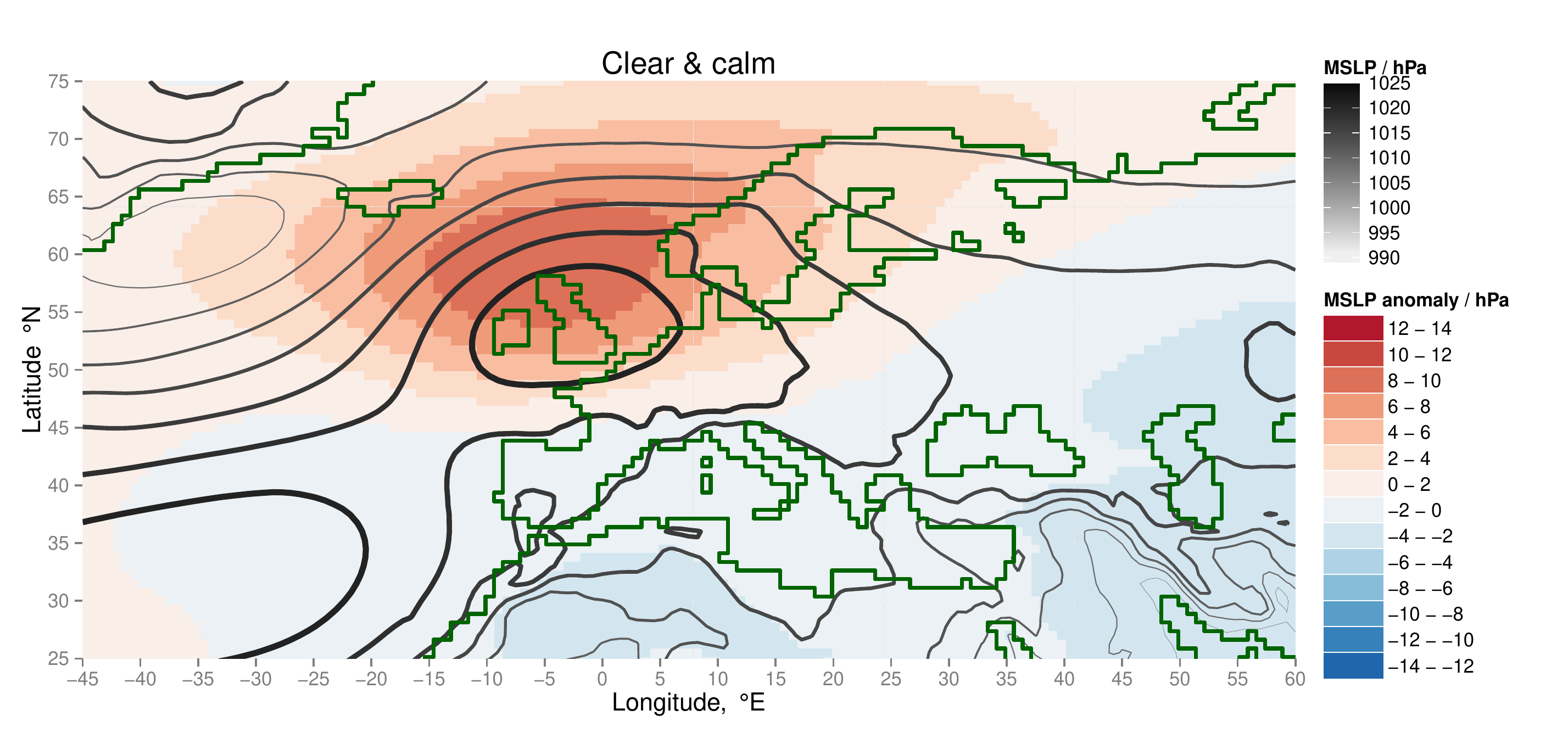} 
\includegraphics[width=0.49\textwidth]{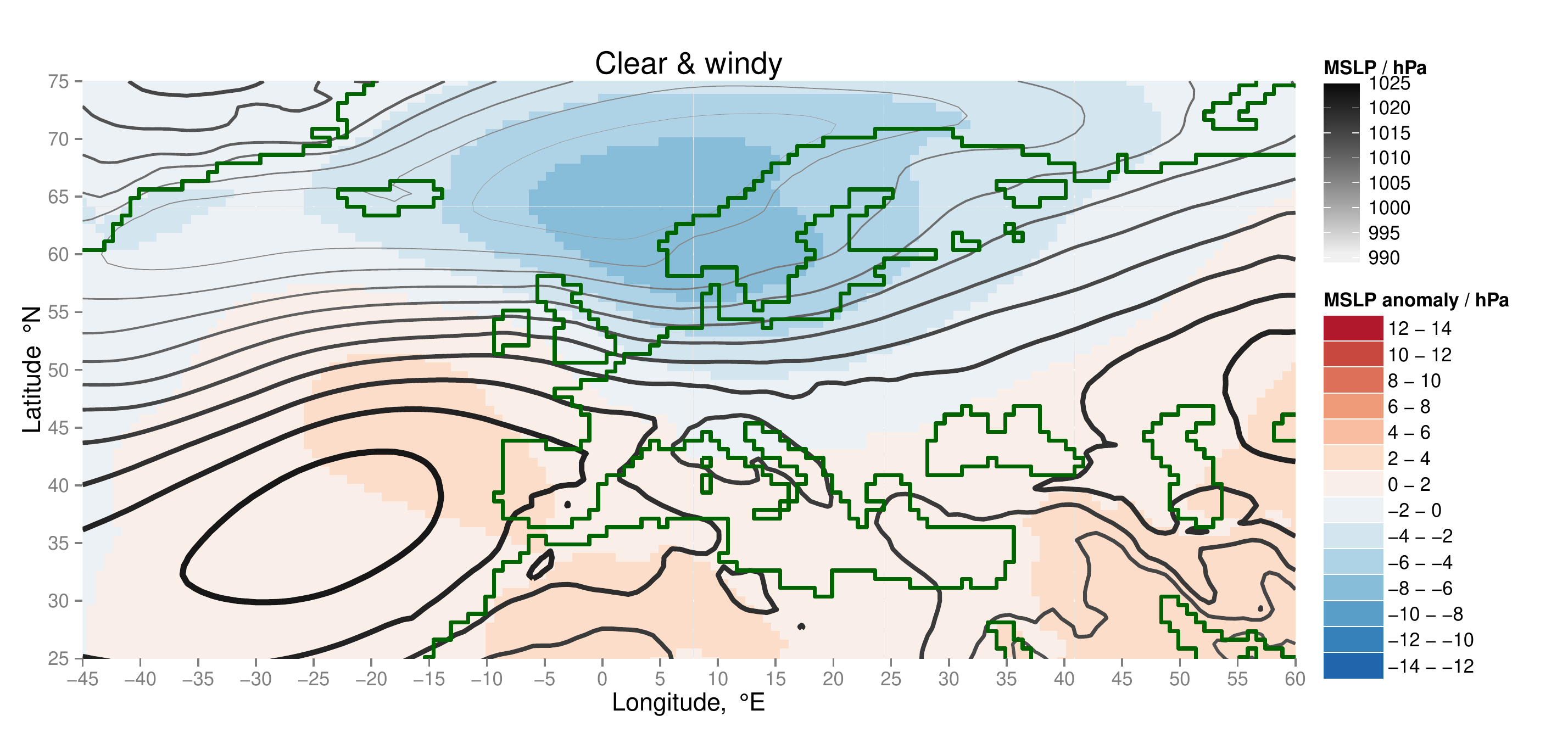} 
\includegraphics[width=0.49\textwidth]{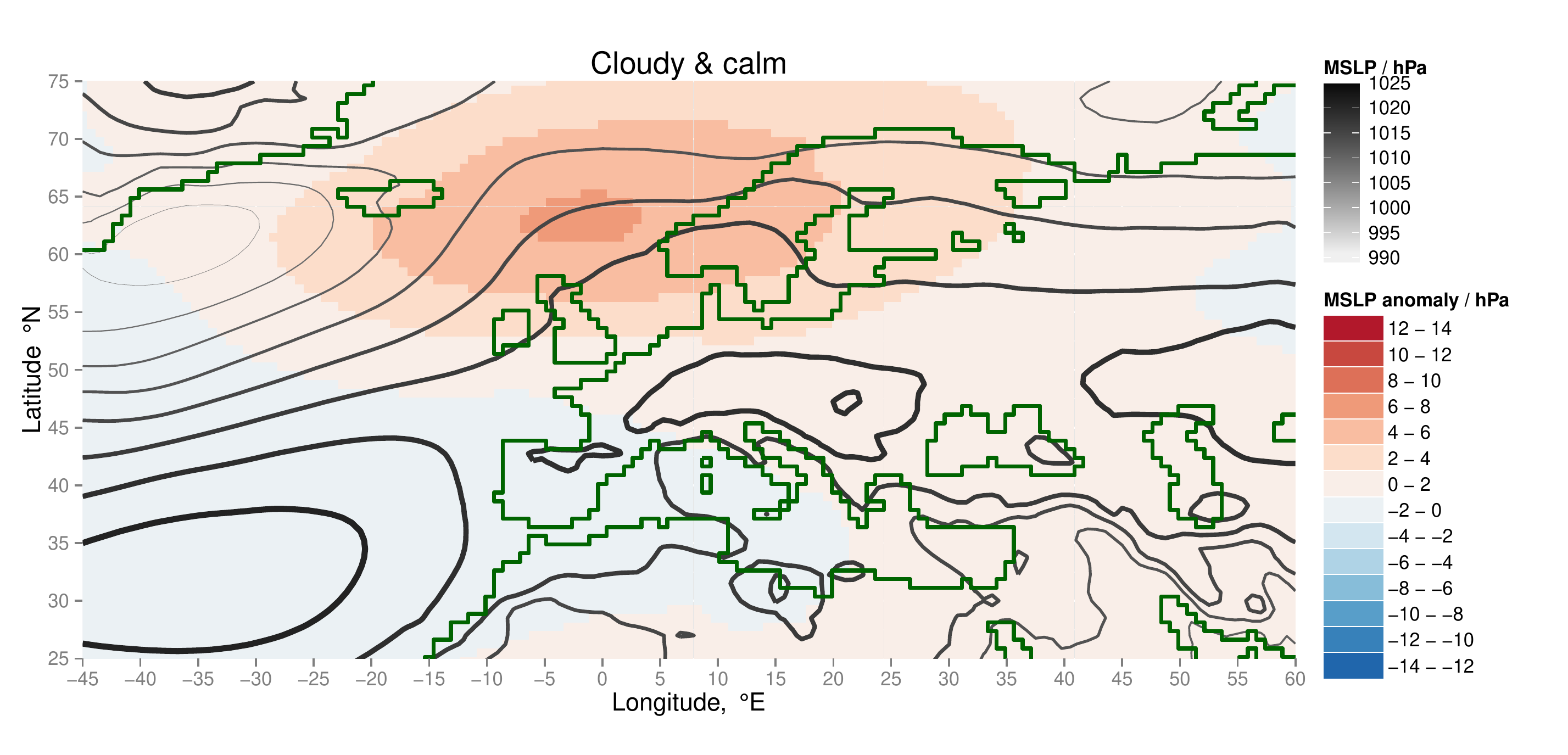} 
\includegraphics[width=0.49\textwidth]{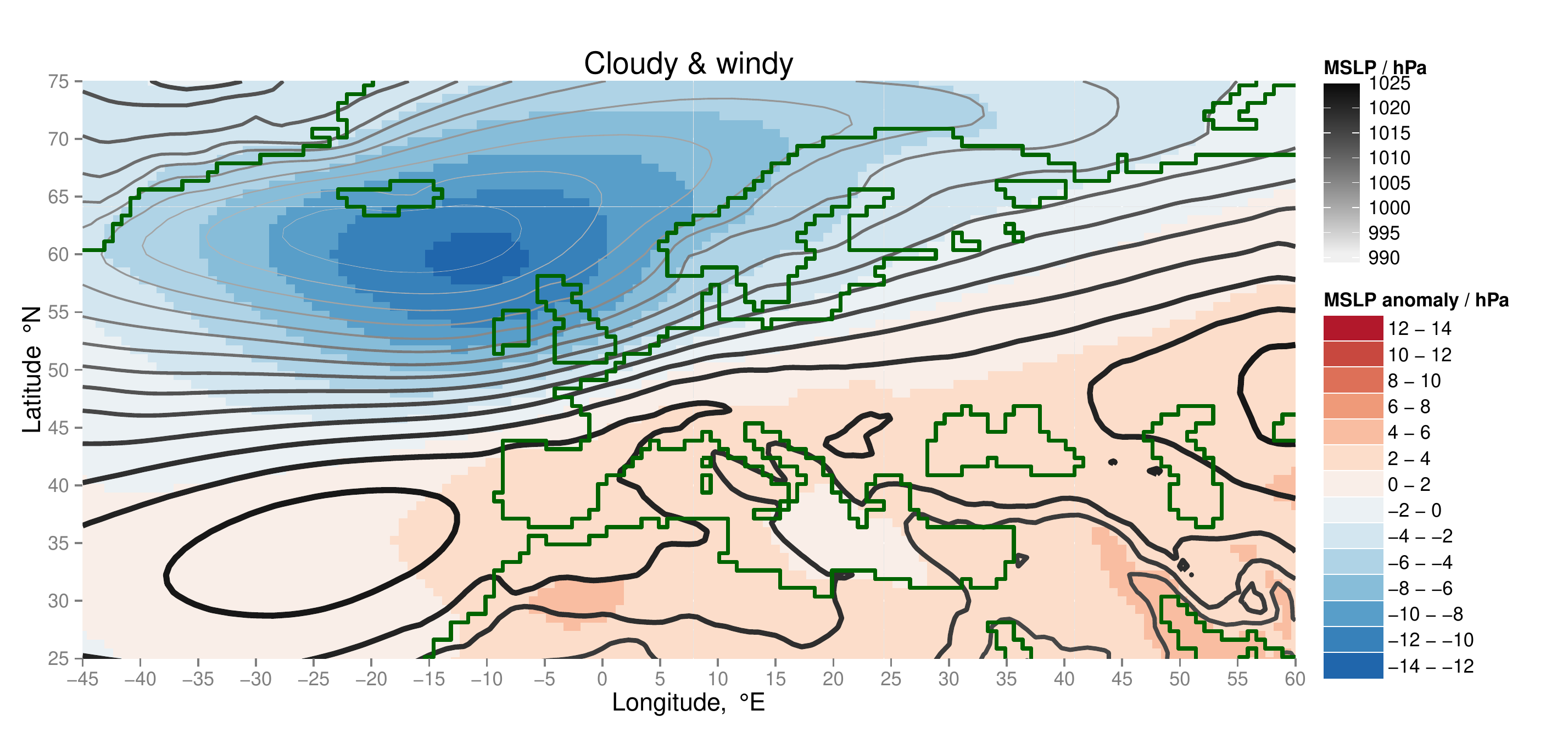} 
\caption{Maps of mean sea level pressure (MSLP) for the four tercile combinations, based on the GB-average data.  The panels are arranged to correspond to the $\sclearness$--$U$ scatter plots. 
The contours show the pressure levels (thicker lines correspond to higher pressure), and the shading gives the pressure anomalies from the all-year long-term average.
}
\label{f:pmslmaps}
\end{figure*}

While we have attributed the regional changes in (anti-) correlation over the year to different terciles in the joint distributions of wind and clearness, we have not considered the large-scale atmospheric features that give rise to such events.  While we reserve a detailed study of this as a topic for future research, considerable insight can be gained by mapping the mean sea-level pressure (MSLP) fields averaged over days selected by the four tercile combinations (Figure~\ref{f:pmslmaps}). The ``clear \& calm'' case shows high pressure over the British Isles, as is typical for this kind of weather.  The ``cloudy \& calm''  case looks very similar, but without the high pressure centre over Britain; weak frontal systems would still be able to pass over the country.  The ``cloudy \& windy'' case looks like typical stormy winter weather, with the tight gradient between high and low pressure centres sending strong winds to Britain from the west.  Finally, the ``clear \& windy'' case is similar to the ``cloudy \& windy'' case, but with the low pressure centre weakened and positioned further east, and the high pressure extending further north.  It would be interesting in future studies to consider how many of these clear/windy cases occur soon after a cloudy/windy situation.

Ultimately, it is important to remember that we are looking at relatively weak tendencies, nudging the distribution of daily-mean events towards or away from the general weak anticorrelation of wind and clearness.  The low resolution of ERA-Interim also means that many relatively small-scale features and processes will not be adequately captured in the data.  On the east coast of Britain, these include land/sea breezes and North Sea fogs for example, which are  important aspects of the local climatology \citep{Mayes2013Regional, Wheeler2013Regional}.

\subsection{Irradiance under high and low wind conditions}\label{s:hilo}
A key application of the joint distributions  we have described is in  understanding the distribution of irradiances under particularly low and high wind conditions; that is, understanding the range of possible solar power output under situations of GB-wide low and high wind energy supply.

Since the operating thresholds for wind turbines do not vary with season, it is important that we do the same, picking thresholds to define ``low'' and ``high'' wind speeds using all data throughout the year.  We use the 10th and 90th percentiles of GB-averaged wind speeds ($3.95\,\windunit$ and $10.9\,\windunit$ respectively) from our ERA-Interim data.

\begin{figure*}
\centering
\includegraphics[width=\textwidth]{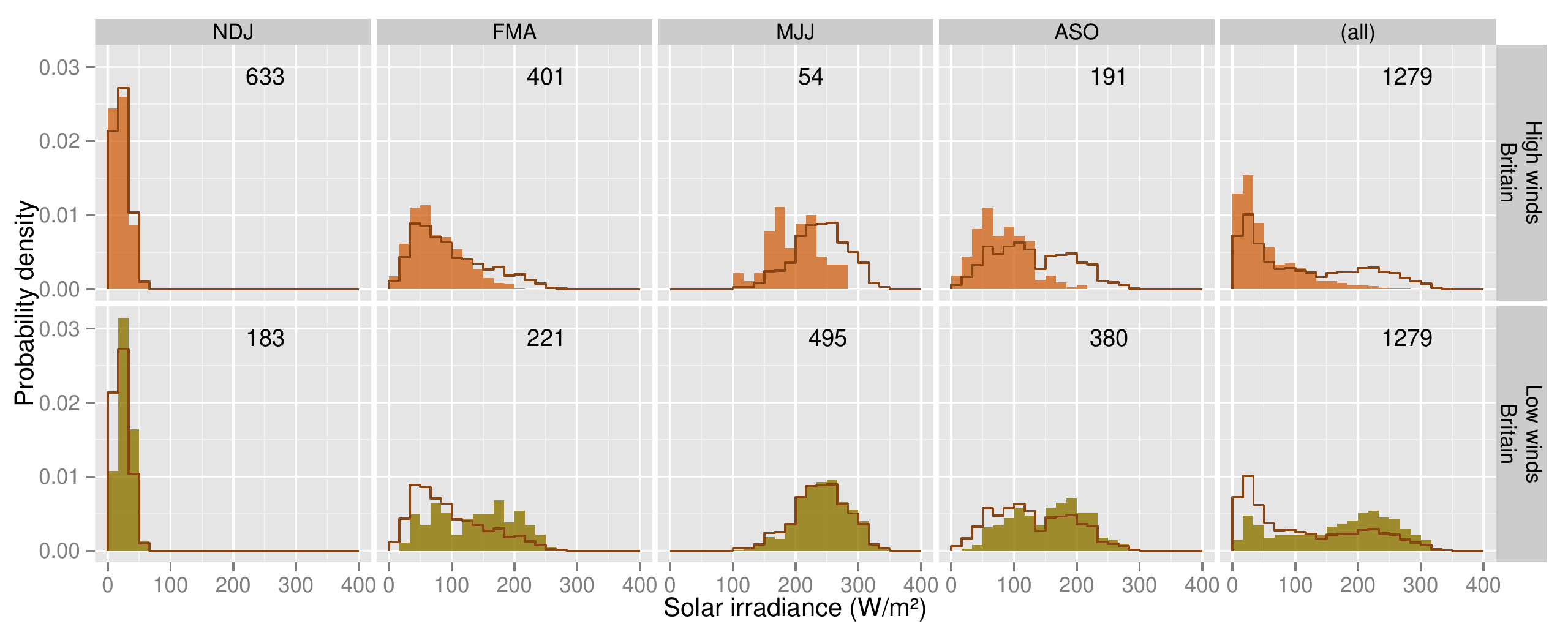}
\caption{Distributions of daily-mean irradiance for Britain.  Each column shows a different season, as labelled.  The outline histograms are for all days in the given season, and the filled histograms show the distribution when selecting only high-wind or low-wind days (upper and lower rows, as labelled).  The number of low/high-wind days in each case is shown in each panel.
}
\label{f:solpdfsgb}
\end{figure*}

The irradiance distributions for high and low wind days in different seasons, and throughout the year, are shown in Figure~\ref{f:solpdfsgb}.  (Note that this is only a particular view of data already presented in the joint distribution in Figure~\ref{f:jointdistro}.)  When considering the all-year case, it is clear that the irradiance distribution for high-wind days is skewed darker, and the low-wind days are skewed brighter.  As already discussed, the seasonal cycle dominates variation in the daily irradiance distribution.  However, Figure~\ref{f:solpdfsgb} shows that this shifting of the distribution holds to some extent in all seasons.

It is important to note that, as has been seen in Figure~\ref{f:jointdistro}, there are very few high-wind days in the summer, and relatively few very calm days in winter. In these cases, the resulting distributions are more noisy. Similarly, the winter high-winds and summer low-winds cases contain such a high proportion of the available days in those seasons that the irradiance distributions do not have the freedom to show a shift relative to their all-winds case. These sampling problems make it more difficult to use the affected distributions to infer probabilities for future events; a longer climatological sample would be preferred.

\begin{figure*}
\centering
\includegraphics[width=\textwidth]{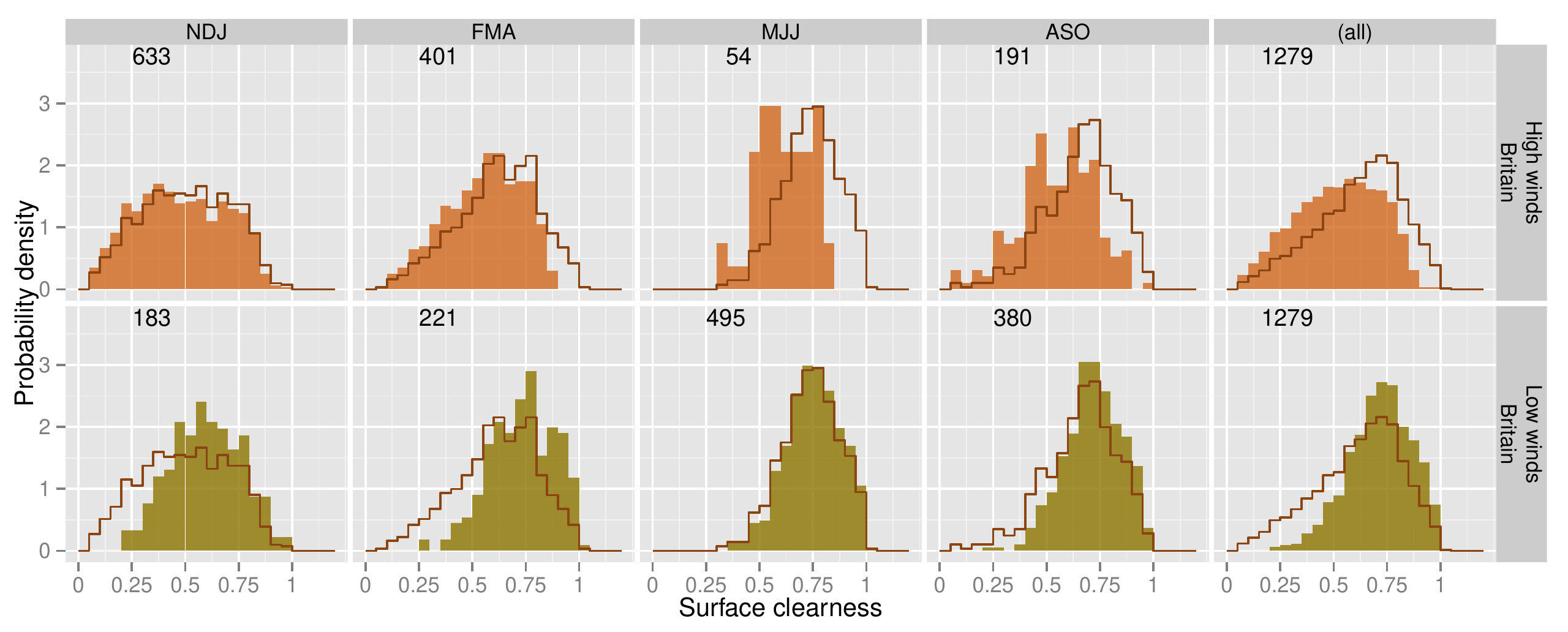}
\caption{As Figure~\ref{f:solpdfsgb}, but using the distributions of daily-mean surface clearness for Britain.  Again, the outline histograms show  all days in the  season indicated, and the filled histograms show the result of selecting high-wind/low-wind days, as indicated.  We still use the `solar' seasons here, for ease of comparison with the previous figure; using the `meteorological' seasons instead (not shown) results in only minor quantitative differences; the qualitative forms of the distributions are unchanged.
}
\label{f:solpdfsgbsolratio}
\end{figure*}

As in  previous sections, we can plot the irradiance in terms of the surface clearness, to remove the direct influence of obliquity in the seasonal cycle, leaving variations due to cloudiness.  This is shown in Figure~\ref{f:solpdfsgbsolratio}, and confirms that low-wind days are preferentially less cloudy (with greater potential for PV electricity generation), and high-wind days are preferentially more cloudy.

While this overall tendency in the distributions is clear (and not unexpected), it is also extremely important to note that the shift in the distributions is relatively small; a significant amount of variability exists even when selecting the extremes of the wind speed distribution.  It is \emph{not} simply the case that \emph{all} windy days are cloudy, nor that \emph{all} calm days are sunny.


\section{Energy balancing}\label{s:balancing}

In this section we consider a key impact of the form of the wind--irradiance joint distribution:  the degree to which power from wind turbines and solar PV panels can complement each other, to reduce energy supply variability on a day-to-day basis over the year.

\subsection{Further methodology} 

We cannot compare the `actual' power output from wind turbines and solar PV panels without a model for the distribution and capacity of such devices across Britain.  While it is feasible to construct such a model that reflects the generation capacities from different sources at a given snapshot in time, renewable capacity in Britain is increasing sufficiently rapidly  \citep{dukes2014ch6, energytrendsMar2015renewables}  that its results would soon be invalid \citep{Drew2015Impact}.  As stated earlier, we aim instead to describe the underlying meteorological aspects of energy supply balancing using historical climatological data, such that the results remain broadly true for many years.

%

Capacity factors are traditionally used when comparing power output from different devices, by taking the ratio of power generated to a standard value defined for each model of wind turbine or PV panel.   In this way, the output from different devices can be compared on an equal basis, removing dependencies on features of particular devices such as their size or efficiency.  However, standard capacity factors from turbines and from PV panels are defined following different principles,\footnote{For wind turbines, a capacity factor would be obtained by scaling by its \emph{rated} power level, $P_U / \Pr$; for PV panels, the power is conventionally scaled by the output under \emph{standard test conditions}, $P_G/P_\STC$ -- see~\ref{s:powercalcs} for details.  These two scales refer to very different physical conditions, and the resulting capacity factors cannot be compared.} and cannot be directly compared; we need a different way of standardising power output.   For this study, we have chosen to scale our estimates of power output $P$ by their long-term mean values $\langle P\rangle$.  (The same approach is taken\footnote{An alternative suggested by \cite{Widen2015Variability} is to scale $P_G$ by the power output under clear sky conditions, similar to what we did in Section~\ref{s:results}. This resulting ``capacity factor'' would, like that of wind speeds, be limited to $[0,1]$; however, it would also remove the genuine seasonal variability in the power output, which is a feature we are interested in assessing here.} by \citealt{Heide2010Seasonal, Heide2011Reduced}.)

To consider the net variability resulting from a combination of wind and solar PV, we should specify their relative capacities with respect to a prescribed total.  We write the total power as
\begin{equation}\label{e:ptot}
 \Ptot 
 = \lambda_U \langle\Ptot\rangle \frac{P_U }{\langle P_U\rangle} 
 + \lambda_G \langle\Ptot\rangle \frac{P_G}{\langle P_G\rangle},
\end{equation}
where the balancing fractions $\lambda_U+\lambda_G = 1$ and the long-term mean total power $\langle\Ptot\rangle$ is a constant to be specified.  \cite{Hoicka2011Solar} and \cite{Liu2013Analysis} used a similar scheme, with an arbitrary total capacity, but assumed equal capacities of wind and solar power.

So, for example, a system with twice as much average onshore wind generation as solar PV would be specified by $\lambda_U=\frac{2}{3}$, $\lambda_G = \frac{1}{3}$.  Solar PV capacity in Britain has been increasing rapidly compared to onshore wind: provisional figures for 2014 show that solar PV comprised 38.6\% of the total PV + onshore wind capacity, and 17.7\% of the total generation from PV + onshore wind \citep{energytrendsMar2015renewables}. UK Government policy in 2014 was for solar PV capacity to reach $10$--$20\,\GW$ by 2020 (\citealt{decc2014solar2}, following \citealt{decc2012roadmap} and \citealt{natgrid2012,natgrid2013}); government estimates of onshore wind capacity, including projects already in the planning pipeline, are for about $16\,\GW$ \citep{decc2013roadmap}\footnote{The plan for \emph{offshore} wind capacity was also to increase to $16\,\GW$ by 2020, and up to $39\,\GW$ by 2030 \citep{decc2013roadmap}.}   Given these rapid developments in renewable energy supply in Britain, we consider the impact of a \emph{range} of different balancing fractions.

In reality, the solar or wind capacity in any given location is subject to important financial and planning constraints, at both local and national levels.  Furthermore,  there is no requirement for wind and solar power to be `well-balanced' against each other; the electricity network is designed to match \emph{demand} rather than produce a constant supply, and incorporates a diverse range of energy sources beyond just onshore wind and solar PV.  However, understanding the theoretical ability of wind and solar power to balance each other within a region is helpful to inform planning decisions when designing a future energy system.

Finally, it is important to note that we are calculating this daily total power using the GB-averaged wind and solar data, looking at GB-wide balancing, rather than the GB-average of local balancing.  Wind turbines and solar panels are not usually co-located at present, and we are assuming that the electricity network can freely redistribute  power around the island.  Potential generation from offshore wind turbines is likewise not included; the higher wind speeds offshore, consequent higher variability, and much greater planned capacity, means that we can take for granted that it will dominate over any realistic solar capacity scenario.  It is envisaged that future studies into the impact of weather and climate variability on UK energy supply would use more detailed supply and demand models, incorporating offshore wind, wave energy, and hydropower, as well as interconnections with supplies from continental Europe.

\subsection{Balancing results}  

\begin{figure}
\centering
\includegraphics[width=0.49\textwidth]{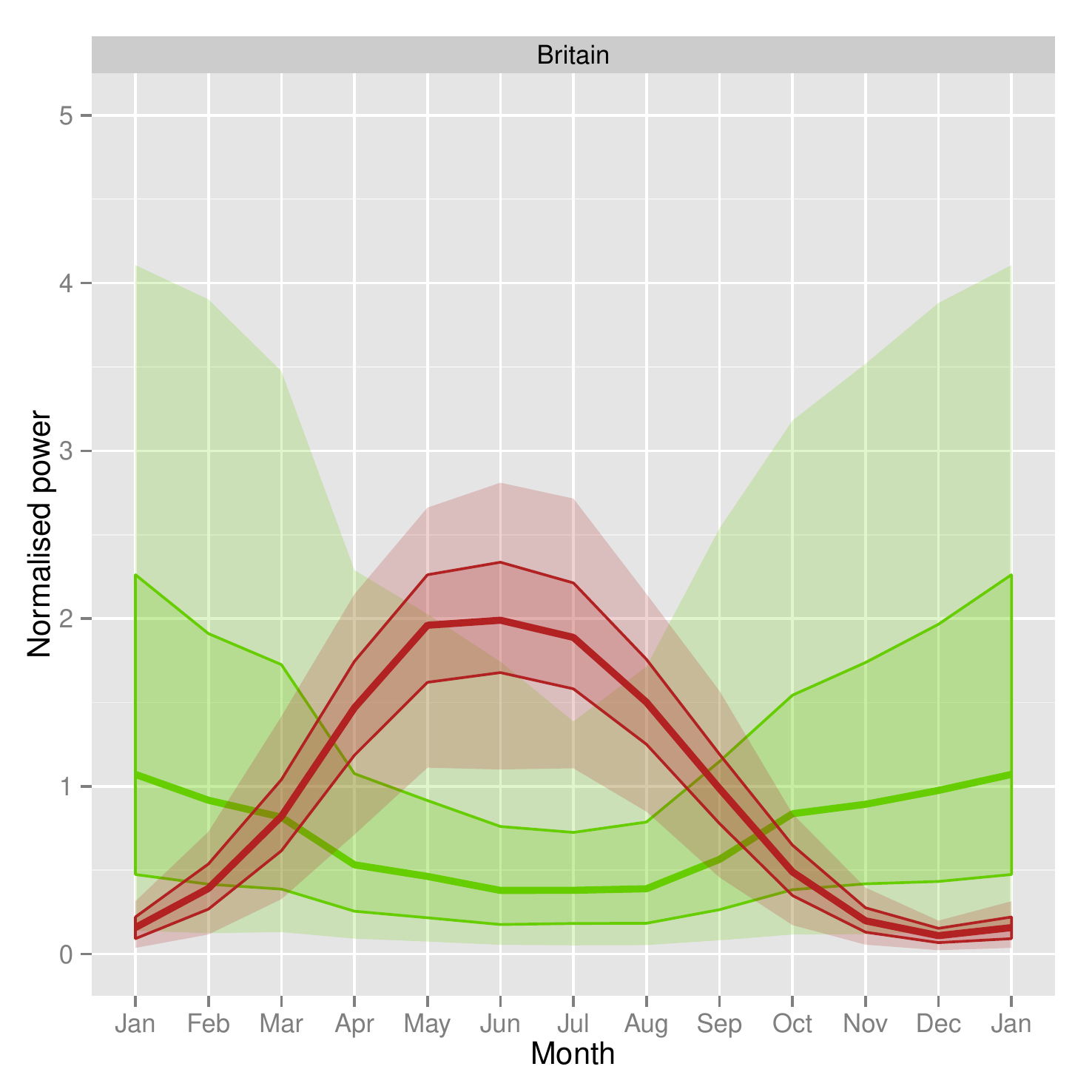} 
\caption{The distribution of daily-mean wind (green) and solar PV (red) power output each month, both scaled by their long-term all-year average (i.e. $P_U/\langle P_U\rangle$ and $P_G/\langle P_G\rangle$ respectively). The lines and shading indicate the medians, 25th \& 75th percentiles, and 5th \& 95th percentiles of the daily data.  Note that, as in previous plots, the results for January are repeated after December to show continuity of the annual cycle.
}
\label{f:balancinggb}
\end{figure}

We show the GB-average monthly distributions of daily wind and solar relative power output in Figure~\ref{f:balancinggb}.   Many of the features of the joint distribution seen in previous sections are visible again here: the strong variability of the wind across all seasons, the larger seasonal variation in solar power, a general anticorrelation between them, and a 1-month lag moving their seasonal cycles away from being exactly in antiphase.

Consider the width and skewness of the distributions in Figure~\ref{f:balancinggb}.  The normalised wind power is very wide, and skewed with a longer tail towards higher values, reflecting the cubic relationship between wind speed and power density.  The normalised solar power on the other hand remains more-or-less symmetric about its median value each month.  The variability of both wind and solar power each month is correlated to its average value. (These results follow from those shown in section~\ref{s:jointdistros}.)

These features have important implications for energy balancing. Firstly, in winter, the range of relative solar power available is small, while wind is highly variable.  Solar power has very little relative  capacity to  counteract low-wind days in winter (i.e. a substantial amount would have to be installed to do so).

Secondly, in the summer, solar power is at its strongest and most variable, while wind power \emph{retains} a large degree of variability.  Indeed, they have very similar levels of relative variability over May--Jun--Jul.   This suggests that increasing the relative capacity of solar PV to compensate for low wind in winter could have the effect of \emph{increasing} the total variability in summer.

\begin{figure}
\centering
\includegraphics[width=\columnwidth]{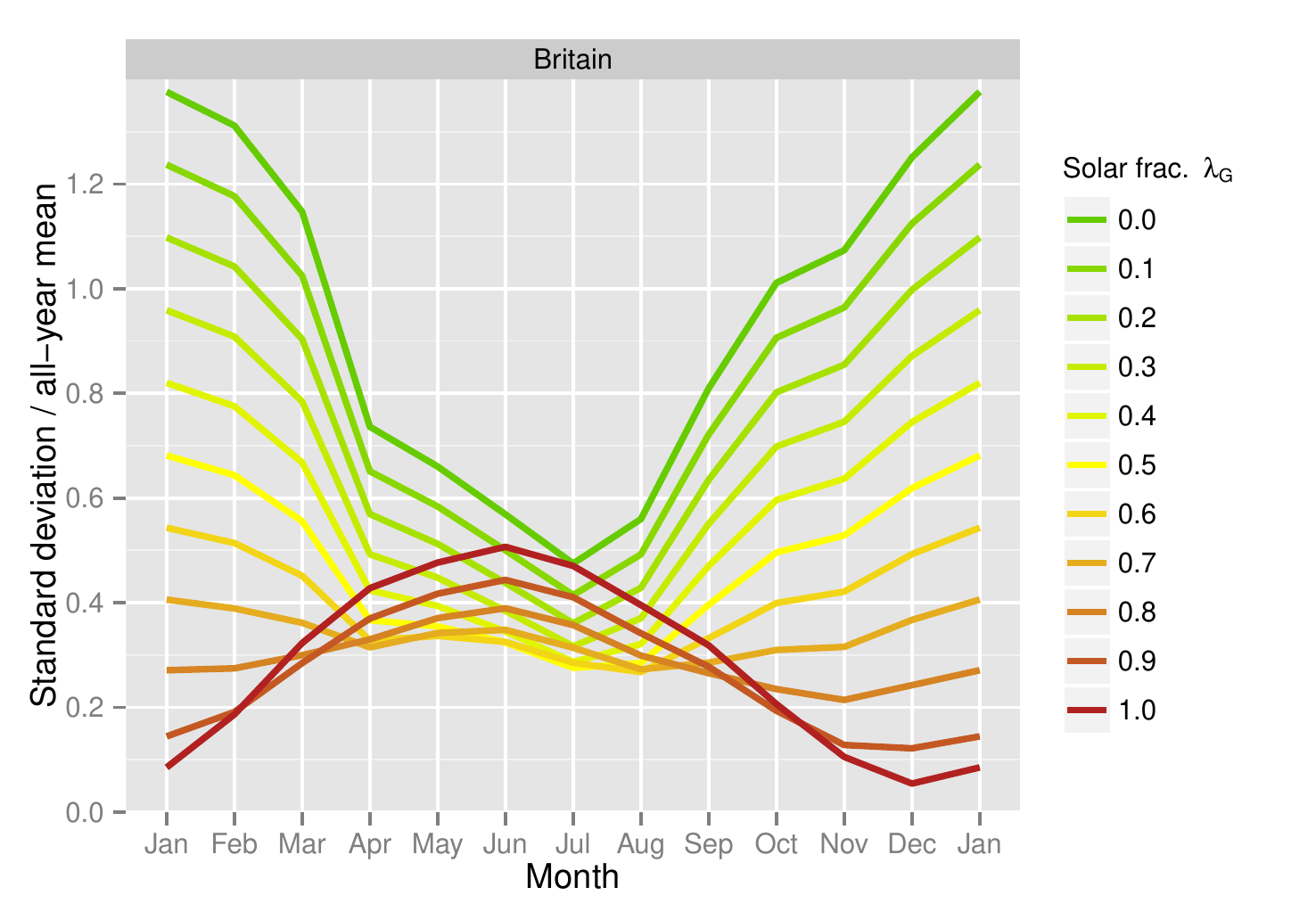}
\caption{Relative variability each month of total power from solar PV and onshore wind ($\Ptot$, see equation~\ref{e:ptot}), in terms of the standard deviation of daily data as a fraction  of the long-term mean (similar to Figure~\ref{f:monsdovermean}).  Each line represents a different balancing scenario, in terms of the fraction of solar capacity compared to the total of wind and solar: the top line in green ($\lambda_G=0$) represents a wind-only scenario, and increasing the solar capacity reduces the relative variability in winter; the lowest line in winter, in red ($\lambda_G=1$) represents a solar-only case.
}
\label{f:balancingvar}
\end{figure}

We now consider directly the following key question: \emph{to what extent is the variability in total power output reduced by incorporating solar PV power?} We show this for different wind/solar balancing scenarios  in Figure~\ref{f:balancingvar}, using the monthly standard deviation of daily $\Ptot$, relative to its overall long-term  mean value $\langle\Ptot\rangle$.

Increasing the relative solar fraction $\lambda_G$ reduces the relative variability in $\Ptot$ in the winter.  The seasonal variability over the whole year is most reduced with $\lambda_G\simeq 0.7$ ($\lambda_U\simeq 0.3$). For $\lambda_G$ larger than this, the variability in summer becomes greater than in winter.  However, there is no scenario that results in higher net relative variability in any month than the no-solar case ($\lambda_G=0$).\footnote{Note that this is a climatological statement from pooling all days over 1979--2013. There will be considerable interannual variability in the energy balancing for any particular month.}  As already mentioned, these results should be taken as indicators of hypothetical outcomes given the meteorology, rather than representing plausible scenarios or recommended, optimal choices.

Again, these results are values for \emph{average} power over the whole land area of Britain. The result from considering the variability in  GB-\emph{total} power output, after modelling the supply network -- with different amounts of installed capacity in different regions, and including offshore wind power, etc -- could in principle be quite different (e.g. \cite{Drew2015Impact}; although the Monte Carlo analysis of \cite{Monforti2014Assessing} using data in Italy for 2005 suggests the impacts of geographical energy balancing could be quite small).  Similarly,  previous studies have suggested that using hourly rather than daily data are likely to be different again \citep{Heide2011Reduced, Widen2011Correlations, Hoicka2011Solar, Liu2013Analysis}.

\section{Summary and conclusions}\label{s:concs}
This paper has described key features of the co-variability of wind and irradiance over Britain, and their potential impacts on energy supply balancing. We have  avoided detailed modelling of electricity networks and the evolving distribution of wind and solar farms, focusing instead on climatological information and scenarios of relative capacity, to keep our results general.

We have shown that the GB-averaged daily-mean wind speeds are weakly anticorrelated with daily-mean irradiances, with Pearson correlation values in different months in the range $-0.4 \lesssim \rho \lesssim -0.2$.  The form of their joint distribution, in particular its bimodal nature, is primarily due to the strong seasonal cycle in irradiances. 
The seasonal cycle in wind speeds is weaker, and the change in daily variability each month is as important a feature as the monthly change in  mean wind speed.  
After the effect of the Earth's tilted rotation is factored out (using the surface clearness parameter), the correlation between windiness and cloudiness remains.  We find that wind and clearness have comparable seasonal cycles in variability, and complementary seasonal cycles in mean value.

The wind--clearness anticorrelation is stronger on the north-west and south-west coasts of Britain than on the east coast: these western regions are hit directly by Atlantic storms, causally relating clouds and wind.  In the east, cloudiness is less dominated by synoptic low-pressure systems, and there is a greater variety of wind--cloud/irradiance states. In particular, it is the increase in the relative frequency of clear-but-windy days, especially in winter, that acts to reduce the level of correlation in the east. The mean pressure field for such situations has a high pressure system to the south-west of Britain and a low to the north-east, implying westerly/north-westerly flow across the country. 

In contrast, cloudy \& windy days have a low situated north-west of Britain, with a tighter pressure gradient and stronger south-westerly flow.  Clear \& calm days are associated with a high pressure system centred over the British Isles.

We have explicitly shown the form of the daily-mean irradiance distribution under high and low wind conditions.  In all seasons, selecting high-wind days leads to dimmer average conditions, and low-wind days lead to brighter average conditions.  However, our results highlight how broad the remaining distributions are; the anticorrelation between wind and irradiance is only weak.

We have explored the consequences of the joint irradiance--wind speed distribution on the potential for energy supply balancing between onshore wind turbines and solar PV panels.    Increasing the amount of solar PV capacity relative to onshore wind reduces the overall variability throughout the year, up to a wind:solar capacity ratio of about 70:30. Further increases in relative solar capacity still reduce total relative variability in winter, but increase it in summer -- although in all months the variability remains below that of the wind-only case.  The ability of solar PV to compensate for lulls in wind power in winter has to be balanced against the risk of increasing variability in summer.

A consequence of our results is to show that, even under the current ambitious government plans for solar PV installation, the power supply from onshore renewables will remain much more variable in winter than summer due to the much greater capacity of wind power.  How this variability would be managed would depend on many factors: the nature of the electricity grid; the availability of other electricity sources (nuclear and gas power stations,  interconnections with mainland Europe), as well as potential forms of energy storage; and the structure of the energy market.  More detailed modelling of particular scenarios, including using a spatially-resolved supply model and/or using higher temporal resolution data,  would enable more precise projections of the impact of meteorology on future energy systems.


\section{Acknowledgements}
This work was supported by the Joint UK DECC/Defra Met Office Hadley Centre Climate Programme (GA01101).  
ERA-Interim data was obtained from the ECMWF archive and are used are used under \href{http://apps.ecmwf.int/datasets/data/interim-full-daily/licence/}{license}.
The authors would like to thank James Manners, Debbie Hemming and Karina Williams for helpful discussions.

\bibliographystyle{elsarticle-harv-pbdois}
\bibliography{renewablerels}

\appendix 
\section{Power output calculations}\label{s:powercalcs}
In this appendix we describe how we converted wind speed and irradiance into estimates of power output.

\subsection{Power from wind turbines}
The energy flux -- that is, the power per unit cross-sectional area, i.e. the power (surface) density --  of an air mass of density $\rho$ moving horizontally with speed $U$, is given by 
\begin{equation}\label{eq:windpdens}
  \Pdensu = \frac{1}{2}\rho U^3.
\end{equation}
We will take the air density to be a constant $\rho=1.2\,\densunit$.

\begin{figure}
\centering\includegraphics[width=\columnwidth]{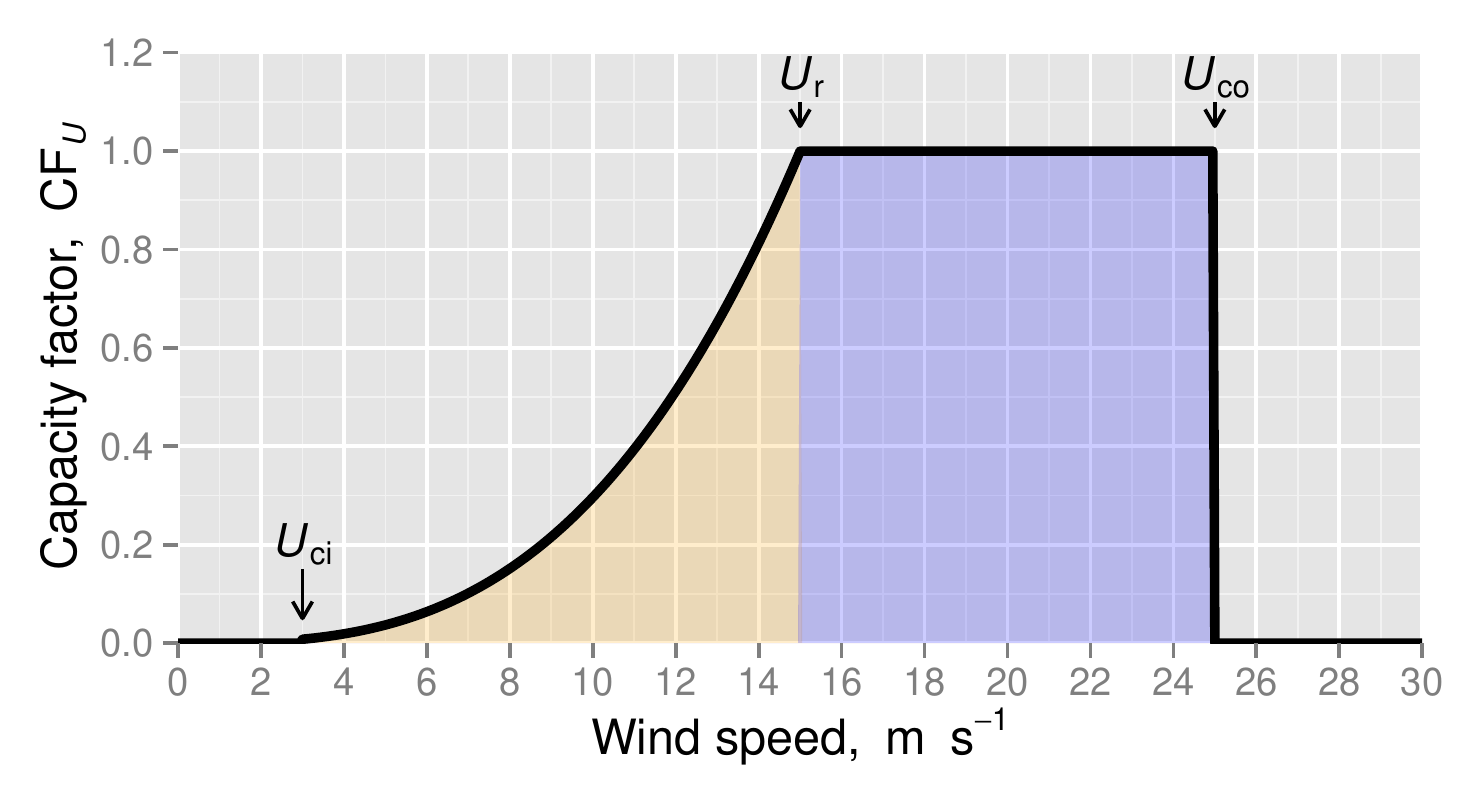} 
\caption{Power curve of the wind following equation~(\ref{eq:windpowerfn}), using the wind speed thresholds given in equation~(\ref{eq:turbinemodel}), shown in terms in the capacity factor $\CFu =  P_U / \Pr$. The ramping region is shaded in orange, and the region with power output at the rated level (equation~\ref{eq:windpowerrated}) is shaded in purple.
}
\label{f:windpowercurve}
\end{figure}

We use a relatively simple model of a wind turbine to determine the full conversion to output power.  This uses a simple power curve (Figure~\ref{f:windpowercurve}), with thresholds describing when the turbine starts generating (`cut-in', $\Uci$), when it is mechanically limited to its peak, or rated, output ($\Ur$), and when it is shut down for safety under high winds (`cut-out', $\Uco$).  The rated power is that given by the power density at wind speed $\Ur$, through the area swept by its blades (of radius $R$), modulo an overall efficiency factor $\eta_U$:
\begin{equation}\label{eq:windpowerrated}
  \Pr = \frac{1}{2}\eta_U \rho \pi R^2 \Ur^3.
\end{equation}
The power output can therefore  be described in full by
\begin{equation}
  \label{eq:windpowerfn}
  P_U = \left\{
    \begin{array}{ll}
      0,                                &           U < \Uci,   \\
      \Pr \left(\frac{U}{\Ur}\right)^3, & \Uci \leq U \leq \Ur,  \\
      \Pr,                              & \Ur  \leq U \leq \Uco, \\
      0,                                &           U > \Uco,    \\
    \end{array}
  \right.
\end{equation}
We use the following values for the wind thresholds \citep[e.g.][]{Brayshaw2011}:
\begin{eqnarray}
  \label{eq:turbinemodel}
  \Uci &=& 3\,\windunit, \\  
  \Ur  &=& 15\,\windunit, \nonumber\\
  \Uco &=& 25\,\windunit. \nonumber 
\end{eqnarray}
In practice, we do not need to specify the efficiency $\eta_U$ or blade radius $R$ if we instead calculate a wind energy capacity factor $\CFu =  P_U / \Pr$.  Doing this minimises the amount of turbine-specific information needed, leaving the results more general.

It is important to note that in this paper we are only using daily-mean wind speeds for this estimate, whereas in reality the wind speed thresholds in the power curve are defined for instantaneous wind speeds.   However, the vast majority of daily-mean wind speeds in our ERA-Interim data occur in the ramping region where $P_U \propto U^3$, which is also the case for wind speeds in `real life'.

The power curve we have described is necessarily only an approximation to the real behaviour of a wind turbine.  All turbines exhibit significant scatter in power output around their nominal power curves \citep[e.g.][]{Kiss2009Comparison}, and many groups have taken different approaches to empirically characterising the wind--power relationship \citep[e.g.][]{Lydia2014Comprehensive}.  The uncertainty due to assuming a constant air density is of a similar magnitude to the scatter around a power curve.

\subsection{Power from solar photovoltaic panels}
The power output from a solar photovoltaic panel depends on both the total incident downwelling irradiance $G$ and the ambient air temperature $T$.  Following \cite{Huld2008Geographical},  we write the power generated from a solar PV panel as 
\begin{equation}\label{eq:solpower}
  P_G = \etarel(G,T) \cdot\eta_\STC\cdot \eta_\mathrm{e}\cdot A\cdot G,
\end{equation}
in terms of the rated module efficiency under ``standard testing conditions'', $\eta_\STC$, the efficiency of other connected equipment (such as inverters) $\eta_\mathrm{e}$, the panel area $A$, and the relative efficiency $\etarel(G,T)$, which captures the environmental dependence of the panel performance.

The ``standard testing conditions'' (STC) refer to an irradiance of $G_\STC=1000\,\Wpersqm$ and a PV module temperature of $T_\STC = 25\,\degC$, at which the PV module generates a power of $P_\STC$.  The STC efficiency is therefore defined as $\eta_\STC = P_\STC / (A\cdot G_\STC)$.

The PV module temperature is empirically related to the air temperature through 
\begin{equation}
  \label{eq:moduletemp}
    \Tmod = T + \left( T_\mathrm{NOCT} - T_0 \right)  \frac{G}{G_0},
\end{equation}
where the  reference values are $T_0 = 20\,\degC$ for the ambient temperature, and $G_0 = 800\,\Wpersqm$ for the irradiance.  The PV module details are encapsulated in terms of a nominal operating cell temperature under these conditions; $T_\mathrm{NOCT} = 48\,\degC$ is often used.  Note that the STC module temperature $T_\STC$ corresponds to an ambient air temperature of $T= -10\,\degC$.

\begin{figure}
\centering\includegraphics[width=\columnwidth]{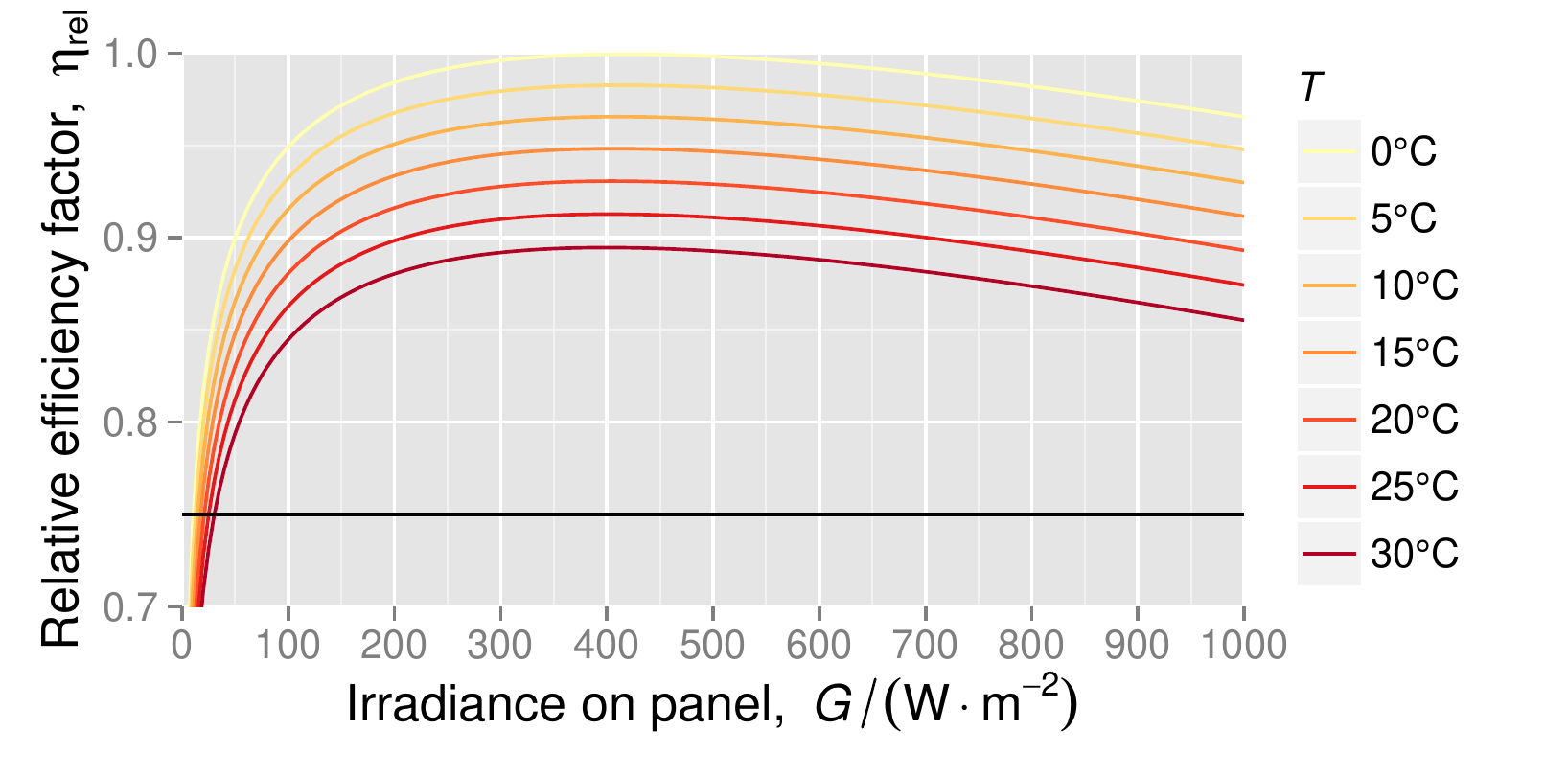}
\caption{Relationship of the solar panel relative efficiency $\etarel$ with incident irradiance and air temperature, following equation~(\ref{eq:solarefficiency}), and the module temperature relationship of equation~(\ref{eq:moduletemp}).   
}
\label{f:etarel}
\end{figure}

The relative efficiency is given by another empirical function:
\begin{eqnarray}
  \label{eq:solarefficiency}
  \etarel(G,T) &=& \left[ 1+\alpha \Delta\Tmod \right] \\
  & & \times \left[ 1 + c_1 \ln G'  + c_2\ln^2 G' + \beta \Delta\Tmod \right],\nonumber
\end{eqnarray}
in terms of the scaled variables $G' = G/G_\STC$ and $\Delta\Tmod = \Tmod - T_\STC$.  The temperatures are in degrees Celsius, and the constants are $\alpha = 4.20 \times 10^{-3}\, \mathrm{K}^{-1}$, $\beta = -4.60 \times 10^{-3}\, \mathrm{K}^{-1}$, $c_1 = 0.033$, and $c_2 = -0.0092$.  Note that under standard test conditions,  $\etarel = 1$ by construction.  The variation of $\etarel$ with air temperature and irradiance is shown in Figure~\ref{f:etarel}.

\begin{figure}
\centering\includegraphics[width=\columnwidth]{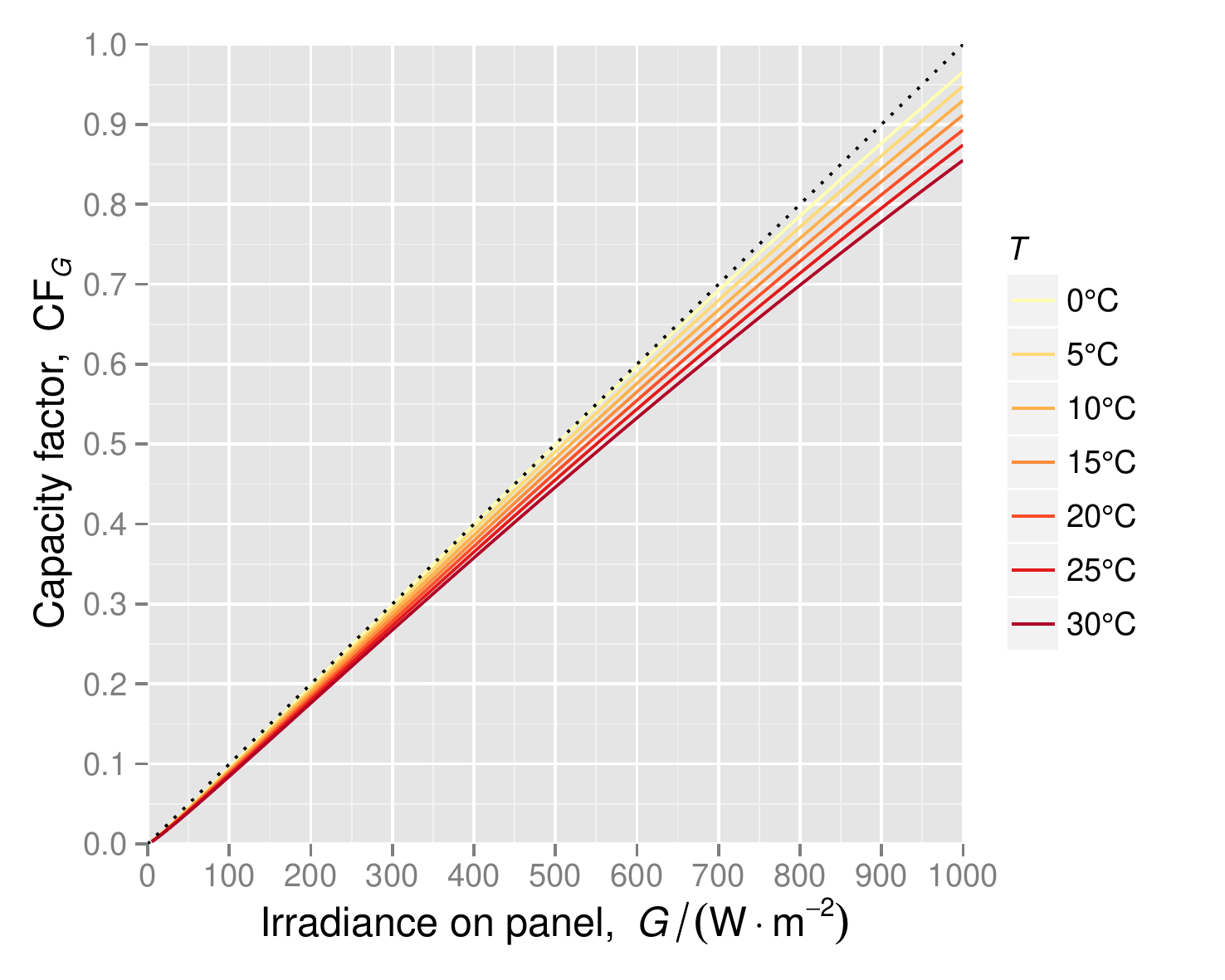}
\caption{Power output from a solar PV panel, relative to the output under standard test conditions (STC, see text for details), following equation~(\ref{eq:solpower}) and the expressions for $\etarel$ and $\Tmod$ given in equations~(\ref{eq:solarefficiency}) and~(\ref{eq:moduletemp}). 
}
\label{f:solarpowercurve}
\end{figure}

As with wind power, we can avoid specifying some PV module details by defining a capacity factor, $\CFg$.  Since the `rated' power for a PV panel is its output under standard test conditions, we have
\begin{equation}\label{eq:solarcapfac}
  \CFg = \frac{P_G}{P_\STC} \equiv \etarel(G,T) \frac{G}{G_\STC}.
\end{equation}
This saves us from specifying the panel area $A$, and the efficiencies $\eta_\STC$ and $\eta_\mathrm{e}$.  The resulting relationship between power output and incident irradiance, at different air temperatures, is shown in Figure~\ref{f:solarpowercurve}.  Note that, unlike in the wind power curve, this model has no upper limit to solar power output, so the capacity factor is not limited to lying between 0 and 1.  While a PV module would in practice have an upper limit to its output, governed by the semiconducting and electronic materials and design, this is likely to be far outside any operational situation seen in practice (indeed, the standard test conditions themselves are outside the range of day-to-day experiences for the UK).

We use these relationships with the ERA-Interim daily mean downwelling shortwave surface irradiance and 2-m temperature fields, allowing us to  calculate an estimate of power generated from a horizontal PV panel.  We are not modelling any spectral response; we are implicitly assuming that this is captured sufficiently by using the standard definition of ``shortwave'' from numerical weather/climate models and satellite systems ($200$--$4000\,\mathrm{nm}$) with the various empirical efficiency factors noted above.  \cite{Dirnberger2015Impact, Dirnberger2015Uncertainty} have shown that the spectral response accounts for an uncertainty of a few percent in the output power.

\section{Joint distributions in terms of power}\label{s:jointdistroalt}
\begin{figure*}
\centering
\includegraphics[width=0.48\textwidth]{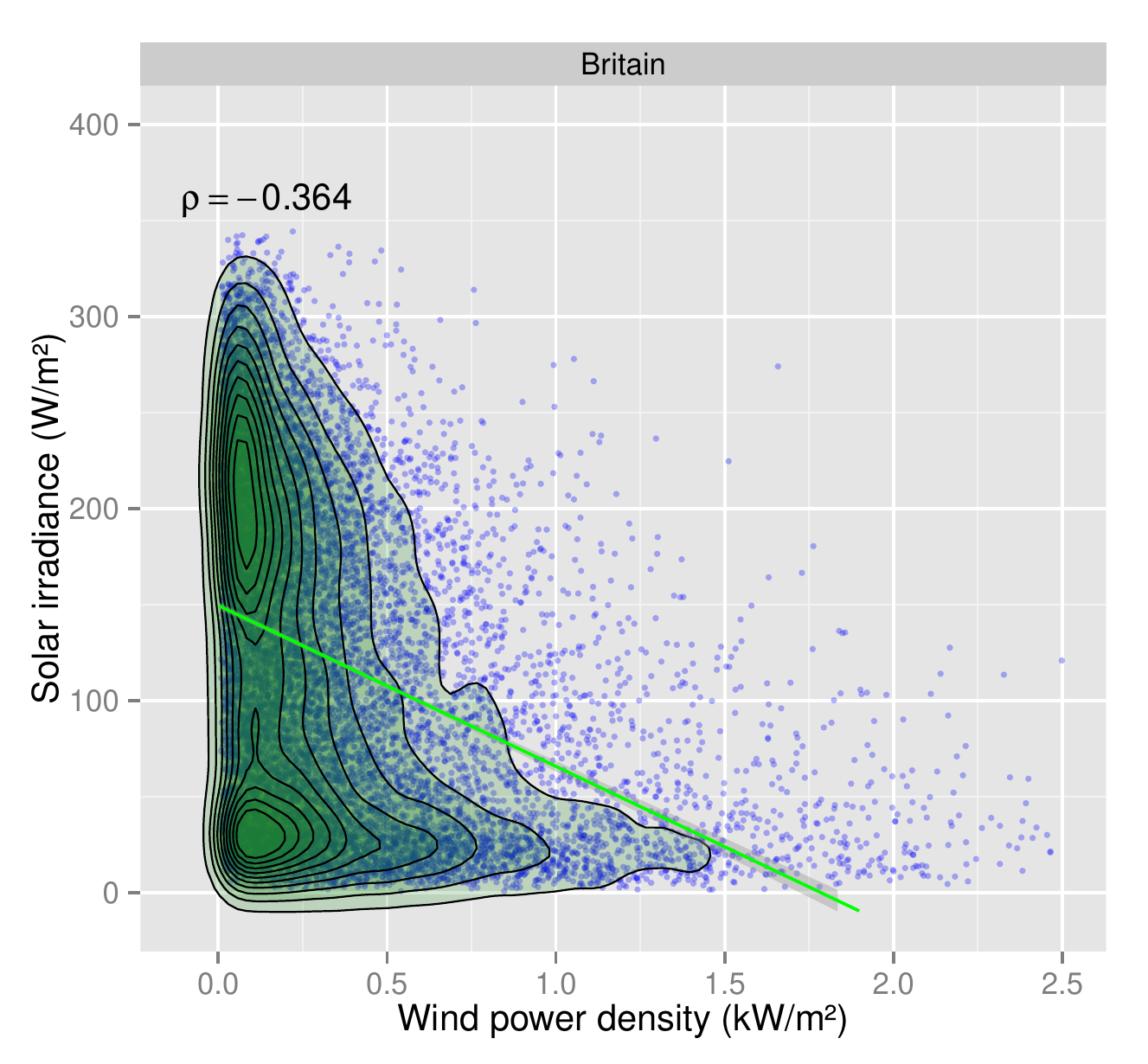} 
\includegraphics[width=0.48\textwidth]{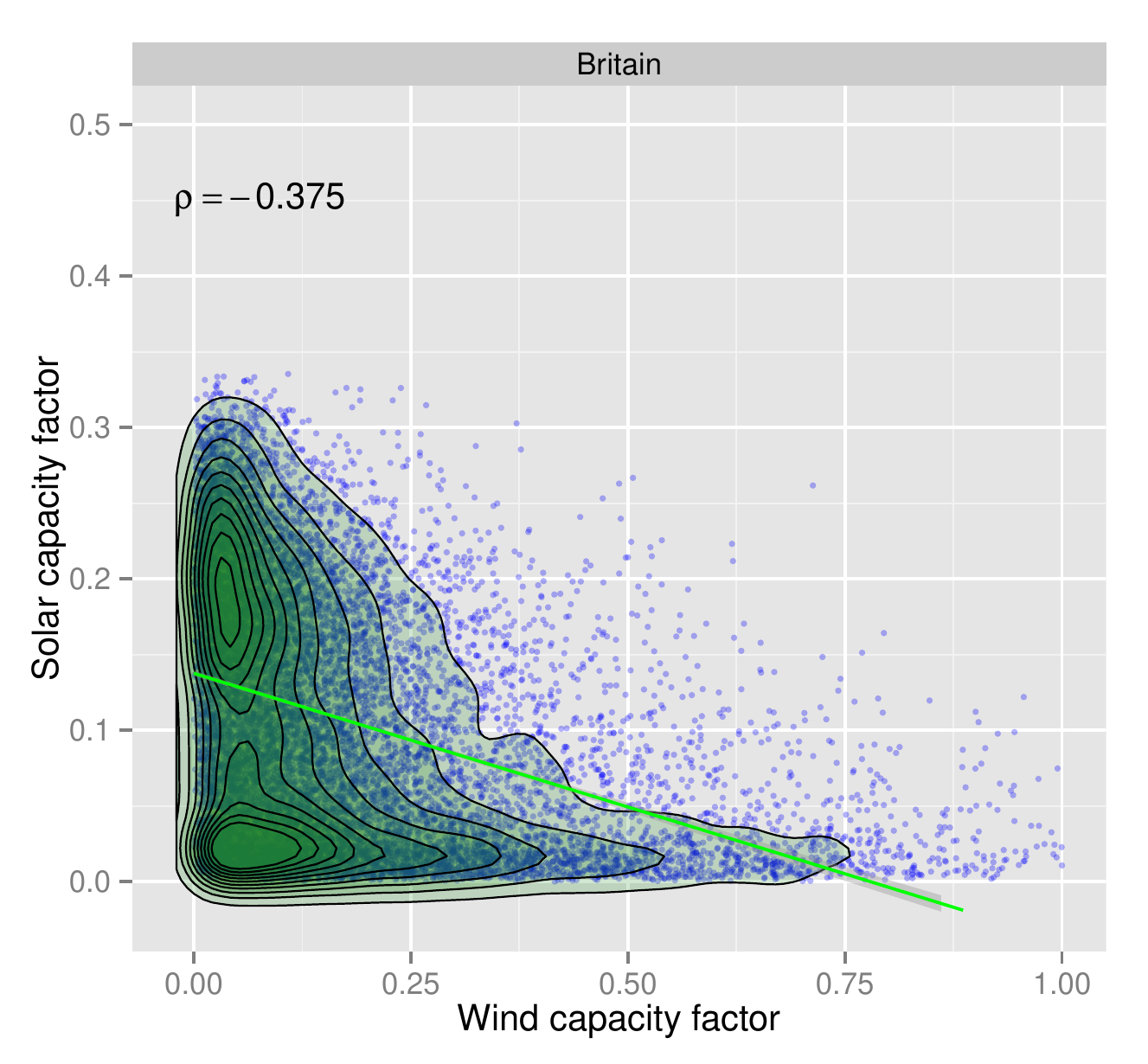}\\
\includegraphics[width=0.48\textwidth]{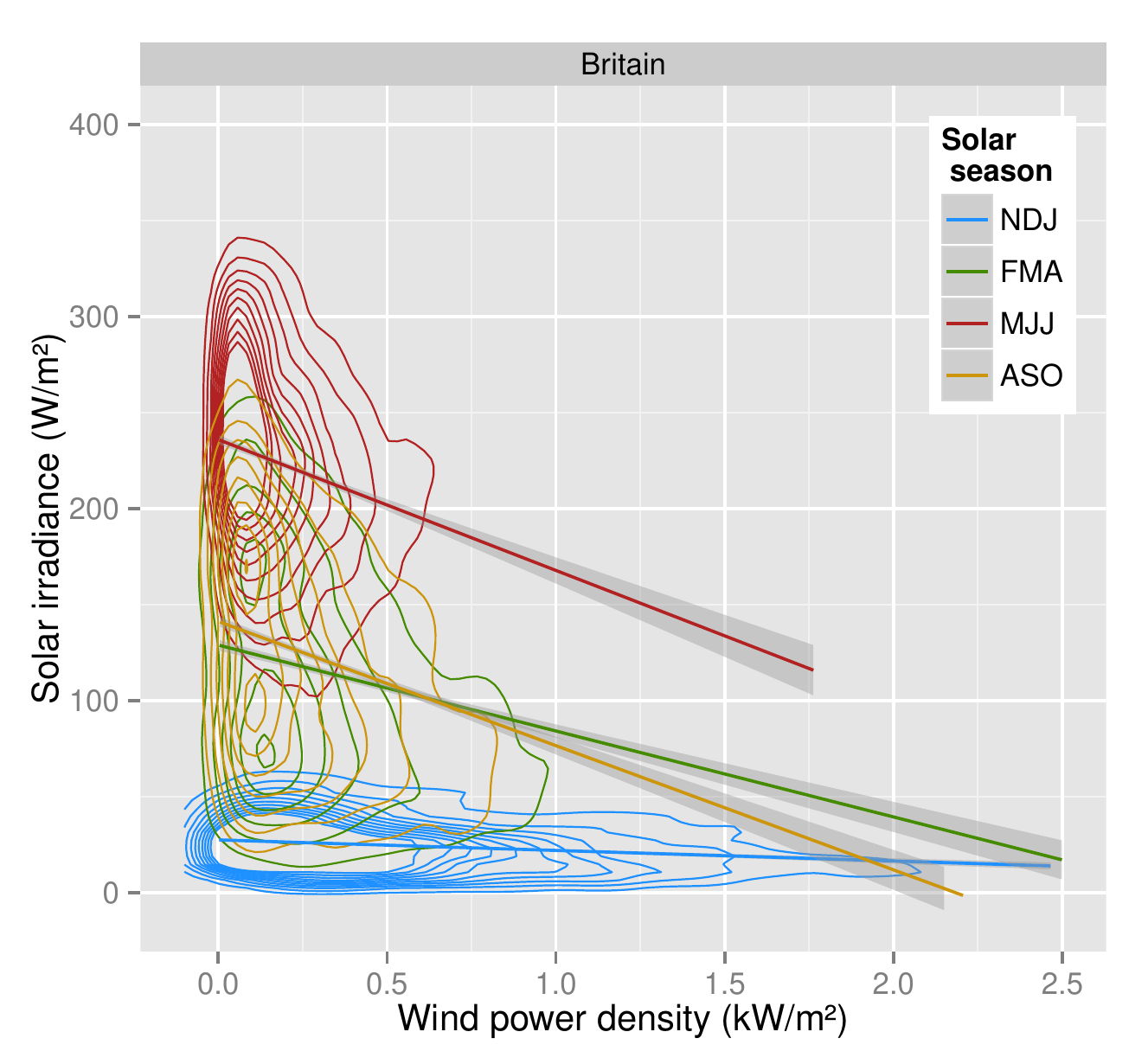} 
\includegraphics[width=0.48\textwidth]{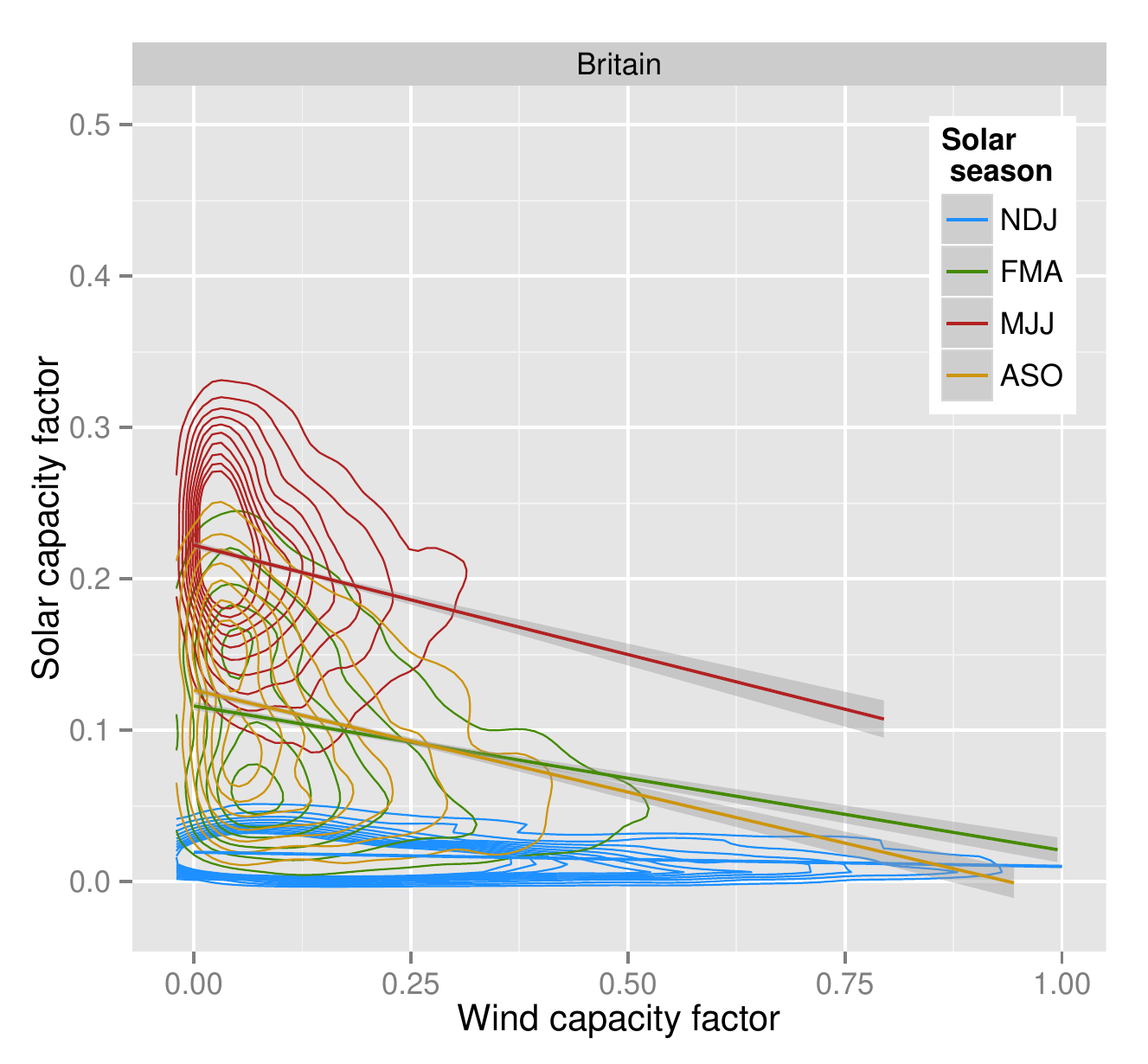}
\caption{Joint distributions of daily-mean data as in Figure~\ref{f:jointdistro}, but using different power metrics for wind and solar PV.  Different contours are used in the four panels.
}
\label{f:jointdistroalt}
\end{figure*}

For completeness, we include in Figure~\ref{f:jointdistroalt} the joint distributions of power density and capacity factor for wind and solar power.  While solar power output is largely proportional to irradiance, the cubed wind speed in the  wind power calculation has a strong effect on the form of the joint distribution.

\section{Spatial variation in wind--irradiance correlations}\label{s:windirradcormaps}
\begin{figure}
\centering 
\includegraphics[width=\columnwidth]{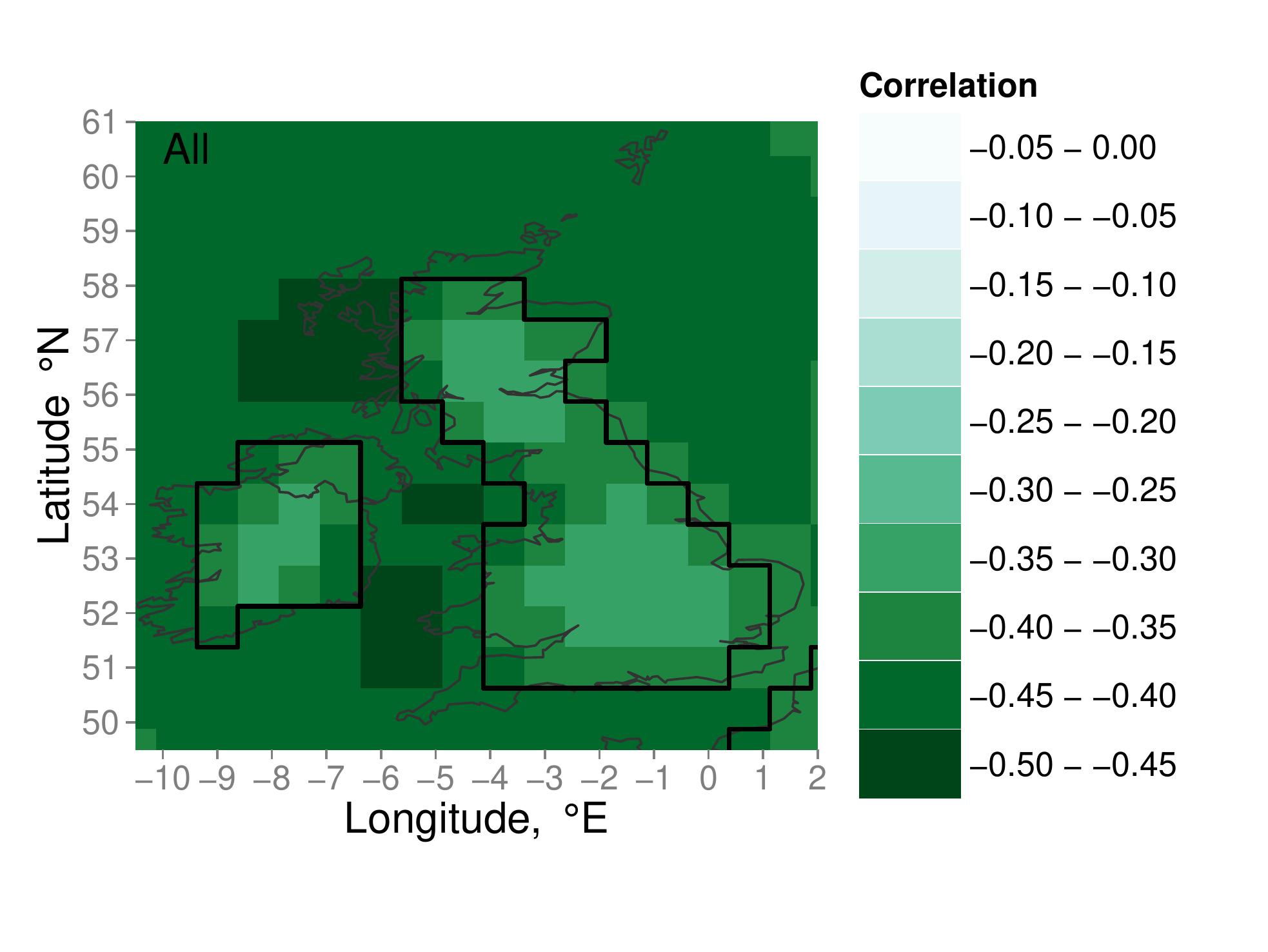}\\
\includegraphics[width=0.48\columnwidth]{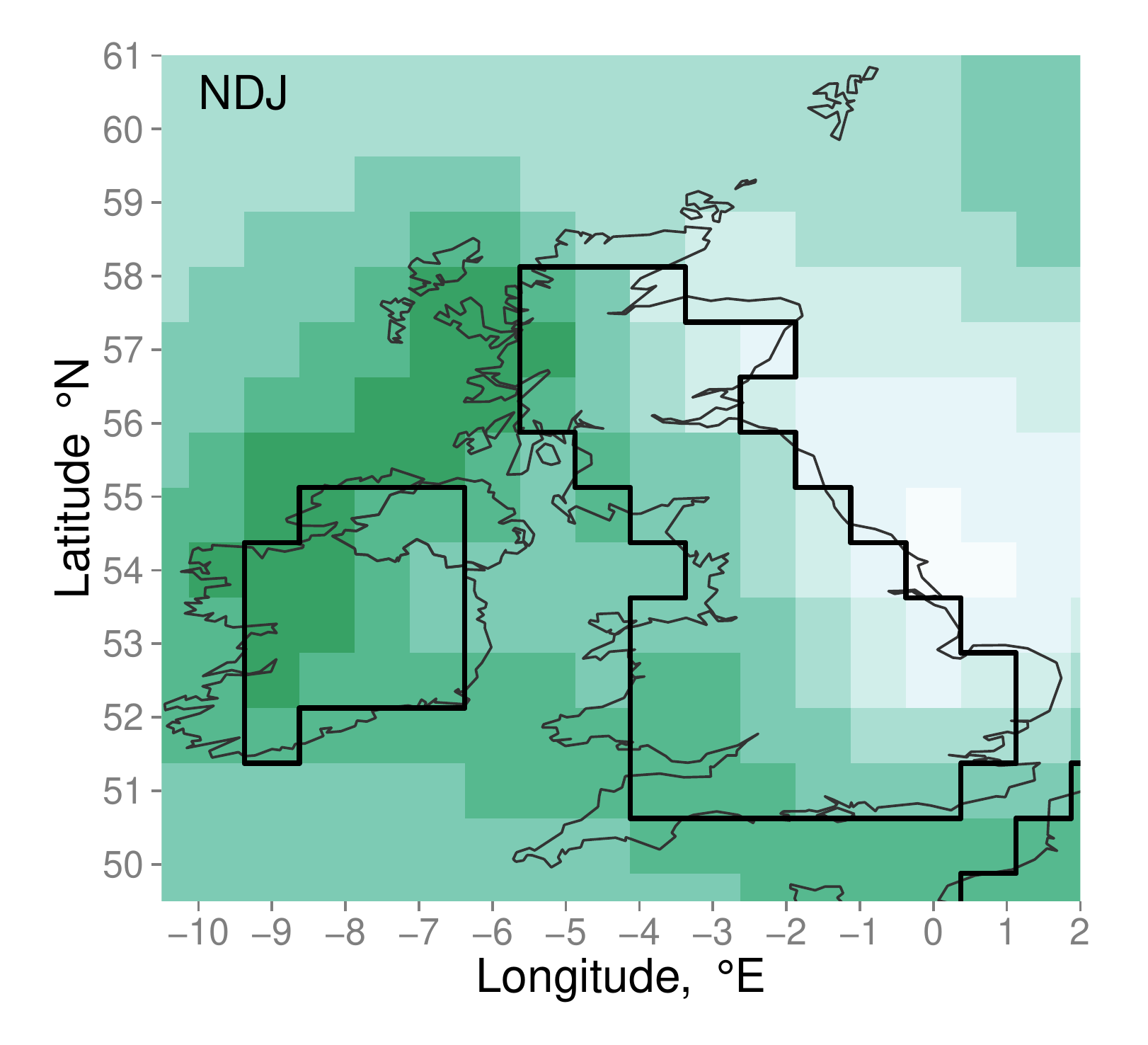}
\includegraphics[width=0.48\columnwidth]{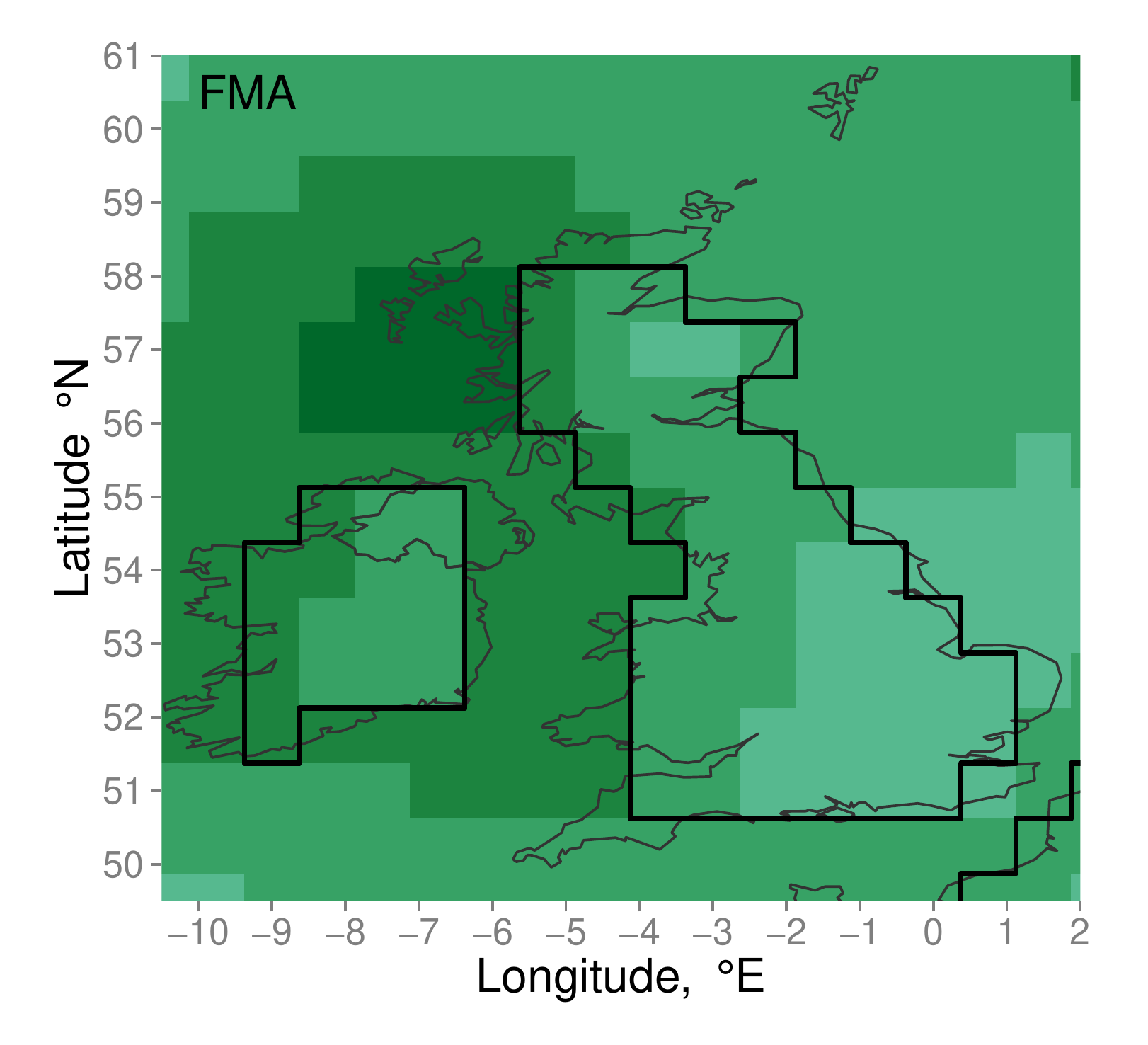}\\
\includegraphics[width=0.48\columnwidth]{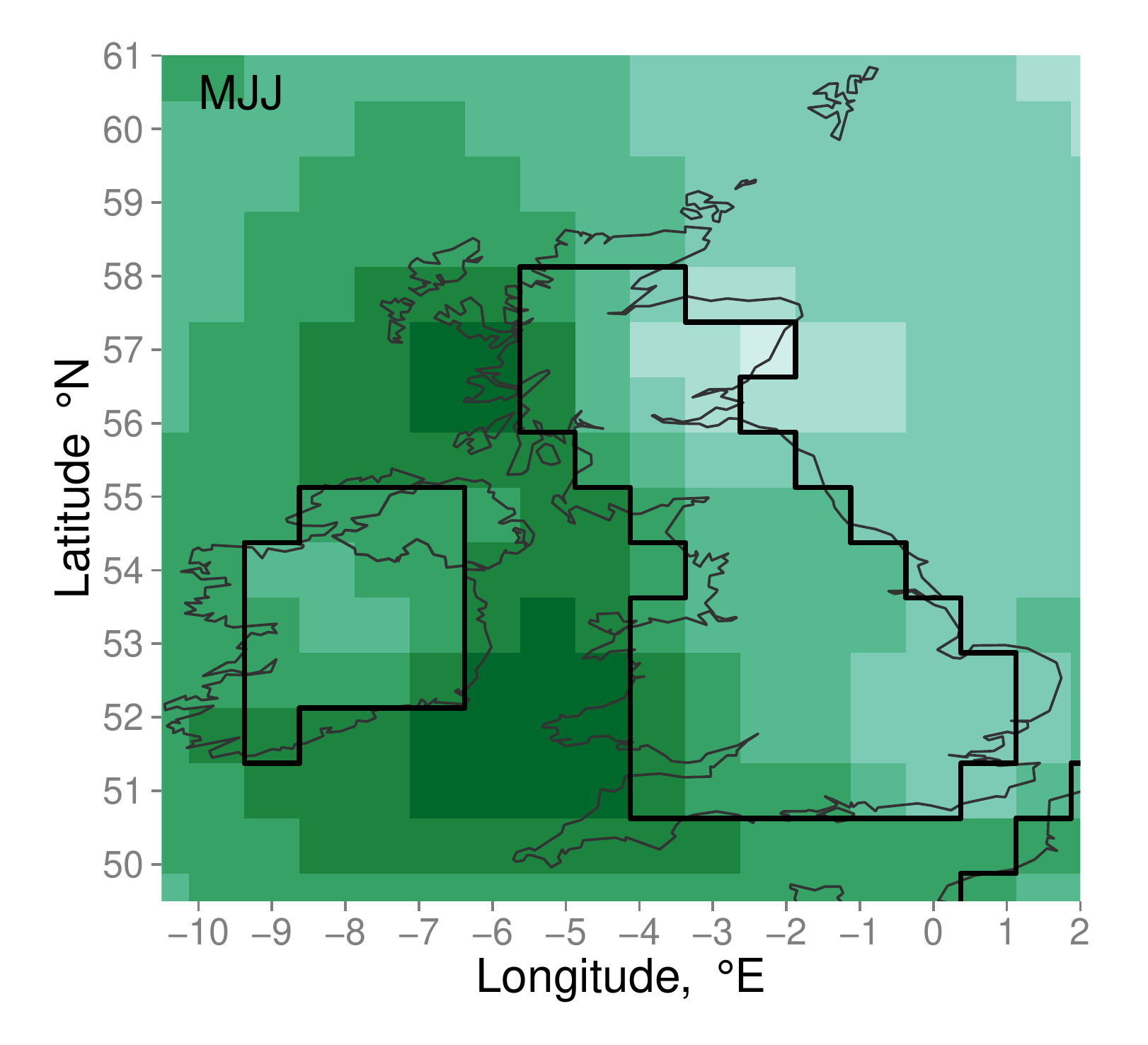}
\includegraphics[width=0.48\columnwidth]{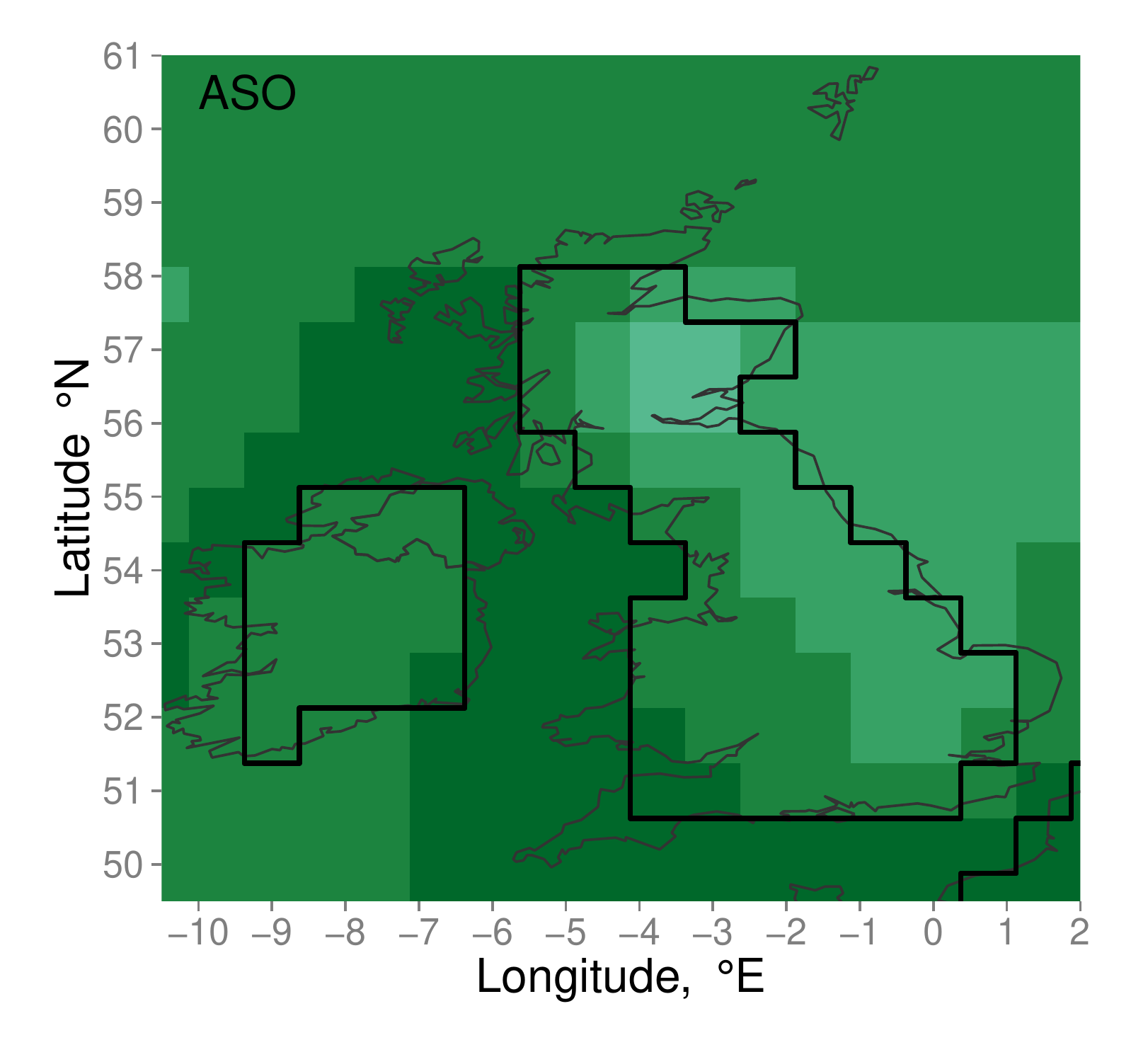}
\caption{Maps of the correlation of daily-mean wind speed with solar irradiance.  The top panels show the all-year correlation, and the smaller panels show different seasons as labelled.  The same colour scale is used in all panels. 
}
\label{f:windirradcorrelmap}
\end{figure}

In section~\ref{s:spatvar} we discussed the spatial variability of the wind--clearness correlation in different seasons, mapping this in Figure~\ref{f:correlmap}.  For completeness, we show in Figure~\ref{f:windirradcorrelmap} the correlation between wind and irradiance, using solar seasons in this case.  The overall picture is similar to the wind--clearness maps, with the strongest anticorrelations off the west coast of Britain, and the weakest anticorrelations off the east coast.

\end{document}